\documentclass[twocolumn]{aastex631}

\shortauthors{Adachi et al.}

\usepackage{graphicx}
\begin{document}

\title{Enhanced gas-phase metallicities and suppressed outflows for galaxies in a rich cluster core at cosmic noon}

\author[0009-0007-9431-6944]{Kota Adachi} 
\affiliation{Astronomical Institute, Tohoku University, 6-3, Aramaki, Aoba, Sendai, Miyagi, 980-8578, Japan}
\email{k.adachi@astr.tohoku.ac.jp} 

\author[0000-0002-2993-1576]{Tadayuki Kodama}
\affiliation{Astronomical Institute, Tohoku University, 6-3, Aramaki, Aoba, Sendai, Miyagi, 980-8578, Japan}

\author[0000-0002-5963-6850]{Jose Manuel P{\'e}rez-Mart{\'i}nez}
\affiliation{Instituto de Astrofísica de Canarias (IAC), E-38205, La Laguna, Tenerife, Spain}
\affiliation{Universidad de La Laguna, Dpto. Astrofísica, E-38206, La Laguna, Tenerife, Spain}

\author[0000-0002-3560-1346]{Tomoko L. Suzuki}
\affiliation{Kavli IPMU (WPI), UTIAS, The University of Tokyo, Kashiwa, Chiba, 277-8583, Japan}
\affiliation{Centre for Data-Driven Discovery, Kavli IPMU (WPI), UTIAS, The Univer-
sity of Tokyo, Kashiwa, Chiba, 277-8583, Japan}

\author[0000-0003-3228-7264]{Masato Onodera}
\affiliation{Department of Astronomical Science, The Graduate University for Advanced Studies, SOKENDAI, 2-21-1 Osawa, Mitaka, Tokyo 181-8588, Japan}
\affiliation{Subaru Telescope, National Astronomical Observatory of Japan, National Institutes of Natural Sciences (NINS), 650 North A’ohoku Place, Hilo, HI 96720, USA}

\begin{abstract}
We present the result of near-infrared spectroscopy using Keck/MOSFIRE for 23 member galaxies in an X-ray cluster XCS2215 ($z=1.46$) to investigate the environmental dependence of gaseous flows and metallicities.
We find that the metallicities derived from H$\alpha$ and [N\,{\footnotesize II}] emission lines of the cluster galaxies are enhanced by 0.08--0.15 dex with $\sim$2 $\sigma$ significance compared to field counterparts for the same stellar mass.
It suggests that inefficient gas accretion in the shock-heated intracluster medium (ICM) in the cluster core results in the lack of metallicity dilution.
We also estimate the mass-loading factor by comparing the observed galaxies with the chemical evolution model that takes into account the outflow processes on the metallicity versus gas mass fraction diagram constructed together with the ALMA data.
We find that the outflows from galaxies in the cluster core region tend to be weaker than those of galaxies in the general field.
It is likely due to the confinement of gas by the high pressure of the surrounding ICM in the cluster core, which leads to the recycling of the outflowing gas that comes back to the system and is used for further star formation, resulting in the progression of chemical evolution.
Compared with higher redshift protocluster galaxies at $z>2$, which tend to show lower metallicity than the field galaxies due probably to dilution of metals by pristine gas inflow, we are seeing the transition of gas accretion mode from efficient cold stream mode to the inefficient hot mode.
\end{abstract}


\section{Introduction} \label{sec:intro}

Past observational studies in the last decade found that the cosmic star formation rate per co-moving volume peaks at $z\sim1.5-2.5$, which is often called “cosmic noon” \citep{madau_cosmic_2014}.
In addition, simulations predict that cold streams of gas along filamentary structures can drive galaxy formation and evolution at high redshifts \citep{dekel_cold_2009}.
Moreover, the contribution of (proto)cluster galaxies to the cosmic star formation rate density is predicted to increase towards high redshifts from only $\sim 1 \mathrm{\,\%}$ in the local Universe to $\sim 20 \mathrm{\,\%}$ at $z\sim2$ \citep{chiang_galaxy_2017}.
Therefore, protoclusters at high redshifts are expected to play important roles in accelerating structure formation and star formation therein, although the coherent picture of the interplay between gas and galaxies across environments and cosmic time is still lacking.

The schematic view of the gas accretion onto cluster haloes is introduced by a numerical simulation \citep{dekel_galaxy_2006} and supported by the later observation \citep{daddi_evidence_2022}.
For the cluster halo with mass of $\log M_\mathrm{halo}/\mathrm{M}_{\odot} > 12$, 
intracluster medium (ICM) is shock heated to a high virial temperature as the cluster potential well grows, and the further gas accretion to galaxies becomes suppressed.
At $z\gtrsim2$, cold streams can still penetrate the hot halo of such massive clusters (``cold in hot" mode), 
while the gas accretion becomes inefficient at lower redshift (``hot" mode).
The transition of these gas accretion phases is predicted to occur at cosmic noon, highlighting the importance of investigating protoclusters at these redshifts.

The connection of galaxy evolution to the surrounding large-scale structure has been investigated in the local Universe by focusing on the environmental dependence of various properties of galaxies, such as star formation rate, color, and morphology \citep{dressler_catalog_1980, lewis_2df_2002, peng_mass_2010}.
These environmental dependencies are thought to originate from differences in gaseous physics around galaxies residing in various environments.
In the cluster environment, it has been suggested that, as a galaxy falls into the cluster center, the gas in the galaxy is stripped by the surrounding ICM (ram-pressure stripping; \citeauthor{gunn_infall_1972} \citeyear{gunn_infall_1972}) and/or tidal effect, which lead to rapid quenching of star formation activity \citep{donnari_quenched_2020}.
In the case of stripping of the gas reservoir in the halo which is loosely bound by a galaxy, star formation activity can continue for a while by consuming the remaining gas within the galaxy but is gradually quenched due to no more gas accretion (``strangulation"; \citeauthor{balogh_ha_2000} \citeyear{balogh_ha_2000}; \citeauthor{larson_evolution_1980} \citeyear{larson_evolution_1980}).
Numerical simulations \citep{dave_galaxy_2011} and observational works \citep{perez-martinez_signs_2023, perez-martinez_enhanced_2024} also suggested that surrounding ICM would push back the outflowing gas driven by active galactic nucleus (AGN) or supernovae (SNe), resulting in lower mass-loading efficiency in the cluster environment (``outflow confinement”).
In addition to these hydrodynamical effects, gravitational effects such as galaxy interactions and mergers can be enhanced in clusters than in the general field and cause rapid quenching followed by AGN or SNe feedback after starbursts.

Such gaseous feeding and feedback processes seen in specific environment also affect gas-phase metallicities in ISM and thus mass-metallicity relation (MZR).
The environmental dependence of MZR for star-forming galaxies is reported from Sloan Digital Sky Survey (SDSS) data \citep{peng_dependence_2014} in the local universe, and gas-phase metallicities are found to have a higher value in galaxies in overdense regions compared to field galaxies.

At higher redshifts ($z\gtrsim 2$), however, the environmental dependence of gas flows and gaseous metallicities are still in debate.
Some works report that galaxies in cluster environments at $1.5 < z < 2.5$ show a higher metallicity than that of the general field \citep{kulas_mass-metallicity_2013, shimakawa_early_2015,maier_cluster_2019,chartab_mosdef_2021,perez-martinez_signs_2023}.
This is interpreted by "strangulation" in which the pristine gas supply is terminated and the gaseous metallicity is no longer diluted \citep{maier_cluster_2019}. Alternatively, the outflow is confined by surrounding ICM, and the chemical enrichment proceeds more efficiently by recycling the polluted gas \citep{shimakawa_early_2015, perez-martinez_signs_2023}.
On the theoretical side, the EAGLE simulation (Evolution and Assembly of GaLaxies and their Environments; \citeauthor{schaye_eagle_2015} \citeyear{schaye_eagle_2015}; \citeauthor{crain_eagle_2015} \citeyear{crain_eagle_2015}), predicts increased metallicity due to the stripping of low-metallicity gas from the outer part of the disks of satellite galaxies at $0<z<2.3$ \citep{bahe_origin_2017, wang_environmental_2023}.

On the other hand, metallicity deficit or little dependence has also been reported for cluster galaxies even at similar redshifts \citep{kacprzak_absence_2015, valentino_metal_2015,namiki_spectroscopic_2019,chartab_mosdef_2021, perez-martinez_enhanced_2024}.
For example, \cite{perez-martinez_enhanced_2024} reported that the low-mass galaxies in $z\sim2.5$ protocluster tend to have higher star formation rates and lower metallicities than field galaxies. This may be due that, at these redshifts, gas accretion is not yet transitioned to the ``hot mode", and the cold, pristine gas can stream into galaxies which dilute the gas-phase metallicities and also enhance their star-forming activities.

Past studies at the cosmic noon have not yet converged to a clear conclusion and show a large scatter among different observations \citep{overzier_realm_2016, perez-martinez_signs_2023, maiolino_re_2019}.
This can be caused by cosmic variance due to a statistically insufficient sample of high-redshift protoclusters, and we still lack a systematic study of the environmental dependence of chemical evolution at these redshifts \citep{maiolino_re_2019}.
In fact, to investigate universal scaling relations such as MZR, the observations toward the lower mass and the detection of relatively faint emission lines are necessary, but it is practically difficult.
The scatter may also reflect the difference in the evolutionary phase of the observed protoclusters. 
The transition of the gas accretion mode can affect the chemical enrichment processes and consequently cause variations in environmental dependence of gaseous metallicity.
There is also no consensus on the physical mechanism of gaseous inflow and outflow and their environmental dependence for cluster galaxies at these high redshifts because of observational difficulties in quantifying these processes \citep{veilleux_cool_2020, davies_jwst_2024}.

In this paper, we focus on an X-ray cluster XMMXCS J2215.9-1738 at $z=1.46$ as our target.
The cluster was first detected by the XMM cluster Survey \citep{stanford_xmm_2006}, and has been intensively investigated by several instruments and over a wide wavelength range from 
X-ray \citep{stanford_xmm_2006, hilton_xmm_2010}, 
optical to near-IR \citep{hilton_xmm_2007,hilton_xmm_2009,stott_xmm_2010,hayashi_high_2010, hayashi_properties_2011,hayashi_mapping_2014,  beifiori_kmos_2017, chan_kmos_2018,maier_cluster_2019},
far-IR \citep{hilton_xmm_2010}, 
and to radio \citep{ma_dusty_2015,stach_alma_2017, hayashi_evolutionary_2017,hayashi_molecular_2018,ikeda_high-resolution_2022, klutse_meerkat_2024}.
This cluster is also one of the targets of the survey called "Mapping HAlpha Line of Oxygen with Subaru" (MAHALO-Subaru; \citeauthor{kodama_mahalo-subaru_2012}\ \citeyear{kodama_mahalo-subaru_2012}), which is the program targeting H$\alpha$, [O\,{\footnotesize III}], and [O\,{\footnotesize II}] emission line galaxies in more than 10 clusters and 2 fields at $0.4 < z < 3.3$.
As a part of this program, \cite{hayashi_high_2010, hayashi_properties_2011, hayashi_mapping_2014} revealed gigantic large-scale structures far out from the cluster core. 
Also, they identified high star formation activities even in the cluster core, unlike in nearby clusters.
Moreover, ALMA observation of the CO(J=2-1) emission line and the derived molecular gas mass mass revealed that there are some gas-rich member galaxies in the cluster core which are recently accreted \citep{hayashi_evolutionary_2017}.
The environmental dependence of the MZR for this cluster is investigated by \cite{maier_cluster_2019} using the VLT/KMOS \textit{H}-band spectroscopy, and the enhanced metallicities ($\sim0.11 \mathrm{\,dex}$) are shown for member galaxies within the half of the virial radius $R_{200}$ compared to the field environment.
The authors suggest that the observed metallicity enhancement can be interpreted by the strangulation scenario, supported by the remaining star-forming activities \citep{hayashi_molecular_2018}.
This cosmic noon cluster is a unique sample due to the large number of confirmed member galaxies by spectroscopy and narrow-band imaging \citep{hayashi_mapping_2014}, along with the available gas mass information necessary for chemical evolution analysis. 
These factors enable us to estimate gaseous outflow rates indirectly by comparing with the chemical evolution models (e.g. \citeauthor{suzuki_dust_2021} \citeyear{suzuki_dust_2021}; \citeauthor{perez-martinez_signs_2023} \citeyear{perez-martinez_signs_2023}; \citeyear{perez-martinez_enhanced_2024}).

This paper proceeds as follows:
Section \ref{sec:data} gives a brief review of our conducted near-IR spectroscopic observations for the cluster galaxies at $z\sim1.5$ and the available archival data.
Section \ref{sec:method} explains the reduction method of the observed data and the derivation of physical quantities from the obtained spectra.
Then, we show results about the star formation activity and metallicity enrichment in the cluster in Section \ref{sec:result}.
We discuss the results with the chemical evolution models in Section \ref{sec:disc}.
Finally, we conclude this analysis in Section \ref{sec:concl}.
In this work, stellar masses and star formation rates (SFR) which are derived assuming the Salpeter IMF in the literature are all converted to the values for Chabrier IMF by dividing by 1.7 \citep{Zahid_2012_census_zo}.
We adopt the concordance $\Lambda$CDM model with $H_0 = 70 \mathrm{\,km \ s^{-1} Mpc^{-1}}$, $\Omega_m=0.3$, $\Omega_{\Lambda}=0.7$.

\section{Data} \label{sec:data}

\subsection{Previous observations for XCS2215} \label{sec:data_past}

As part of the MAHALO-Subaru project \citep{kodama_mahalo-subaru_2012},
a narrow-band survey of [O\,{\footnotesize II}] emission line galaxies in XCS2215 by Subaru/MOIRCS NB912 and NB921 was conducted \citep{hayashi_high_2010, hayashi_properties_2011, hayashi_mapping_2014}.
For 18 central member galaxies, CO(J=2-1) emission lines and dust continua were obtained by past ALMA band-3 and band-7 observations \citep{hayashi_evolutionary_2017,hayashi_molecular_2018}.
Additionally, numerous photometric data in the wavelength range from UV to mid-IR are obtained for these cluster members (See Section \ref{sec:cigale}).
The photometric catalog is constructed by \cite{hayashi_molecular_2018}, and here we briefly describe the data.
In addition to narrow-band NB912 and NB921, Subaru/Suprime-Cam images in \textit{B, Rc, i', z'} band are obtained by \cite{hayashi_high_2010, hayashi_mapping_2014}.
For this cluster region, Canada–France–Hawaii Telescope (CFHT)/MegaCam $u$ and $g$ band from CFHT Legacy Deep Survey and WIRcam $J$, $H$, $K$ band through WIRCam Deep Survey are also available through CFHT Science Archive.
Near-IR images are also obtained by Wide Field Camera 3 (WFC3) on board Hubble Space Telescope (HST) in F125W, F140W, and F160W bands \citep{beifiori_kmos_2017}, 
and by Wide Field Camera (WFCAM) on United Kingdom InfraRed Telescope (UKIRT) in $K$-band \citep{hayashi_properties_2011}.
In the mid-IR range, Spitzer/IRAC provided us with the imaging data in channels [3.6], [4.5], and [5.8] \micron \citep{hilton_xmm_2010}.

We note that the redshifts of galaxies in XCS2215 have been obtained by past optical-NIR spectroscopy \citep{hilton_xmm_2010, hayashi_properties_2011, hayashi_mapping_2014, beifiori_kmos_2017, maier_cluster_2019} and CO(J=2-1) emission line detections with ALMA band-3 observations
\citep{hayashi_evolutionary_2017}.
Also, [O\,{\footnotesize II}] emitters at $z\sim1.46$ can be identified by colour excess of \textit{z} - NB912 (NB921), and their redshifts are determined based on the flux ratios of two adjacent narrow-band filters \citep{hayashi_mapping_2014}.
The measured redshifts range from 1.43 to 1.55, and the cluster membership of each galaxy is identified based on these measurements in this work.

\subsection{Keck/MOSFIRE \textit{J}/\textit{H}-band spectroscopy} \label{observation}

\begin{figure}[tbp]
    \centering
    \includegraphics[width=1\linewidth]{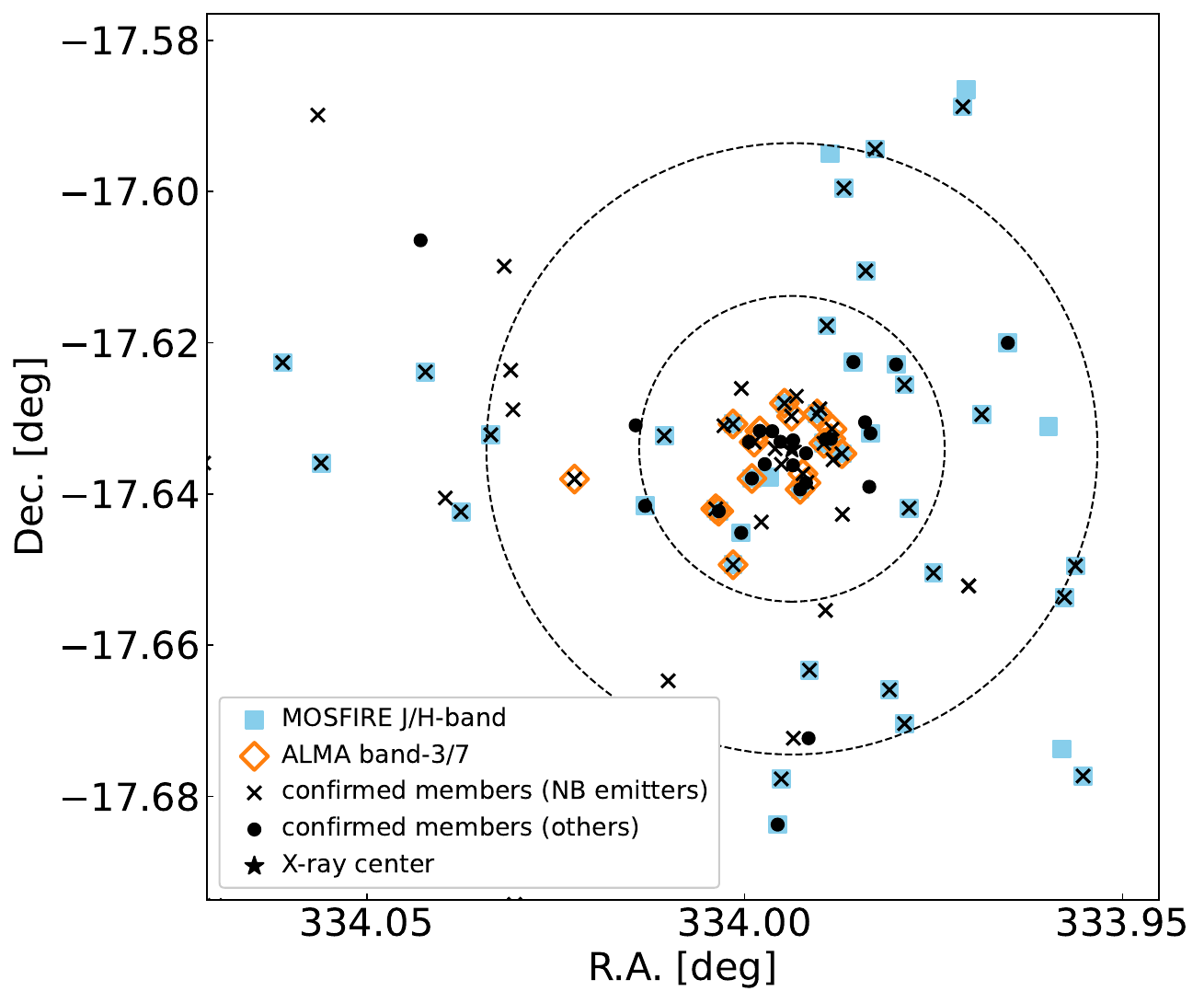}
    \caption{Spatial distribution of MOSFIRE \textit{J}/\textit{H}-band targets in and around XCS2215 cluster (blue squares).
    Orange diamonds represent member galaxies observed by ALMA band-3/7 \citep{hayashi_evolutionary_2017, hayashi_molecular_2018}.
    Among the confirmed cluster members with the determined redshifts, [OII] emitters detected in the NB images are shown as crosses, while the others are shown as points (see Section \ref{sec:data_past}).
    The star mark represents the X-ray center \citep{stanford_xmm_2006}.
    The two dotted circles shows the radii of $R_{200} = 1.23 \mathrm{\,Mpc}$ \citep{maier_cluster_2019} and $0.5\,R_{200}$ from the center, respectively.}
    \label{fig:spatial}
\end{figure}

In order to understand the chemical evolution of galaxies and its environmental dependence at cosmic noon, we conduct NIR spectroscopic observation using the MOSFIRE spectrograph \citep{McLean2010-se, McLean2012-mj} on the Keck I telescope.
To trace SFR and gas-phase metallicities, and classify AGN (see Section\,\ref{sec:line-fitting}) from sources at $z\sim1.5$, 
the spectroscopy was performed using \textit{J}-band (wavelength range of $1.15-1.35$\,\micron; wavelength resolution of $R\sim3318$) covering H$\beta$, [O\,{\footnotesize III}]$\lambda$4959,\,5007 with a single mask, and \textit{H}-band (wavelength range of $1.47-1.80 \ \micron$; wavelength resolution of $R\sim3660$) covering H$\alpha$, [N\,{\footnotesize II}]$\lambda\lambda$6548,\,6584 emission line with two mask configuration.

On June 25 and July 27, 2020, we conducted a MOSFIRE run on Keck telescope (S20A0075N PI:T. Kodama).
We use one mask in \textit{H}-band on the first night and one mask each in \textit{J}-band and \textit{H}-band on the second night.
Taking into account the mask configuration, we select totally 45 objects in \textit{H}-band from the sources located within $R/R_{200} \lesssim 1.5$ as our spectroscopic targets, giving the highest priorities to the 15 objects which have available ALMA band-3/7 data \citep{hayashi_evolutionary_2017, hayashi_molecular_2018} so that we can apply the combined analysis of metallicity and gas mass fraction (Section\,\ref{sec:outflow}).
We also observe 28 member galaxies with the precise redshift measurements in the literature (Section\,\ref{sec:data_past}) as secondary targets.
Finally, we observe two additional objects in the remaining slits, which were previously observed in either NB912 or NB921 but whose redshifts have not been determined.
In \textit{J}-band, we select a total of 18 objects in the same priority scheme, including 13 sources observed in \textit{H}-band.
Figure \ref{fig:spatial} shows the spatial distribution of XCS2215 member galaxies. 
We used ABA'B' dither pattern during the observation, and for each frame, the integration time was 120 seconds.
The net exposure time of the observation on June 25 was 120 minutes in \textit{H}-band, and that on July 27 was 75 minutes each in \textit{J}-band and \textit{H}-band.
5 sources were excluded from the analysis as they were out of slits due to a misalignment of the mask.
The obtained data are reduced using MOSFIRE Data Reduction Pipeline\footnote{\url{https://keck-datareductionpipelines.github.io/MosfireDRP/}} for flat normalization, background subtraction, wavelength calibration, and 1D spectra extraction.
Then, we performed flux calibration on the output 1D spectrum.
Atmospheric absorption was corrected for using a model spectrum of the A0V star and the obtained standard star spectra.
In general, absolute flux values are also calibrated by the observed standard stars.
However, the normalizations of the observed standard star spectra differ between the two slit positions, which cover shorter and longer wavelength ranges.
Therefore, in this work, we use the averaged WIRCam/\textit{H}-band photometry of the targets to convert the count values to fluxes instead of using standard stars.

\section{Method} \label{sec:method}
\subsection{Emission line fitting} \label{sec:line-fitting}

For the reduced spectra, we perform spectral line fitting to obtain emission line fluxes of H$\alpha$, H$\beta$, [O\,{\footnotesize III}]$\lambda\lambda$4959,\,5007, and [N\,{\footnotesize II}]$\lambda\lambda$6548, 6584, which are used for estimation of SFRs and gas-phase metallicities, and classification of AGN.
First, we visually inspect the sources with luminous [O\,{\footnotesize III}] (H$\alpha$) emission lines in the 2D spectra in \textit{J}(\textit{H})-band.
After extracting 1D spectra with MOSFIRE DRP, we mask out the OH line remnants and subtract the fitted continuum for the spectra with sufficiently high ($\mathrm{S/N} >2$) continuum levels.

Here, we conduct a Monte Carlo method to estimate the line flux and its error.
We perform 1000 realizations of the following fitting procedure on each spectrum, adding Gaussian-distributed noise with the observed Poisson noise as the standard deviation to each spectral bin.
We take the median of the posterior distributions as the estimated line flux and take the average of the 16th and 84th percentiles as the 1 $\sigma$ flux error.
The line ratio and its error are also estimated in the same manner.
In a single fitting procedure, the triple Gaussian is simultaneously fitted to the spectra around H$\beta$ and [O\,{\footnotesize III}]$\lambda\lambda$4959,\,5007 (H$\alpha$ and [N\,{\footnotesize II}]$\lambda\lambda$6548, 6584) in \textit{J}(\textit{H})-band, allowing $\pm 5 \mathrm{\,\AA}$ variation from the central wavelength in the observed-frame.
Here, the free parameters in the fitting procedure are the central wavelength, FWHM, and amplitudes of the three emission lines.
Note that we set the same FWHM value for these three lines, assuming that they are all associated with the same H\,{\footnotesize II} regions with the same velocity dispersions.
Moreover, the flux ratios within [O\,{\footnotesize III}] and [N\,{\footnotesize II}] doublets are fixed to satisfy theoretical constraints of $F_{[\text{O\,III}]\lambda 5007}/F_{[\text{O\,III}]\lambda 4959}\sim3$ and $F_{[\text{N\,II}]\lambda 6584}/F_{[\text{N\,II}]\lambda 6548} \sim 3$ \citep{storey_theoretical_2000}.

As a result, H$\alpha$ is detected with $\mathrm{S/N} \geq 2$ for 23 objects out of the 45 MOSFIRE \textit{H}-band targets.
[N\,{\footnotesize II}]$\lambda$6584 and [O\,{\footnotesize III}]$\lambda$5007 in \textit{J}-band are detected for 13 and 4 sources out of them, respectively.
Among the 15 sources observed by ALMA in the \textit{H}-band targets, H$\alpha$ emission line is detected for 5 out of them.
Unspecified luminous emission lines are detected for 2 sources out of those previously observed in either narrow-band filter.
We do not discuss them in this work because their accurate redshifts have not been determined.
We focus on the 23 H$\alpha$-detected galaxies in the following analysis, and other emission lines with $\mathrm{S/N} < 2$ are treated as upper limits in this work. 
The emission line fitting results for these targets are shown in Appendix \ref{sec:app-line}.

We utilize the H$\alpha$ emission line flux to derive SFR for our samples.
We estimate dust attenuation in H$\alpha$, $A(\mathrm{H\alpha})$, assuming the dust attenuation law by \cite{calzetti_dust_2000} and using color excess in stellar emission $E(B-V)_\mathrm{stellar}$ derived from SED fitting (Section \ref{sec:cigale}) : 
\begin{equation}
A(\mathrm{H\alpha}) = k(\mathrm{H\alpha}) \cdot E(B-V)_\mathrm{stellar} / f
\end{equation}
where $f$ is color excess ratio between stellar continuum and nebular emission, defined as $E(B-V)_{\mathrm{stellar}} / E(B-V)_{\mathrm{nebular}} = 0.83$ \citep{kashino_fmos-cosmos_2013}.
We derive SFR for H$\alpha$-detected galaxies via the relation calibrated for nearby H$\alpha$ emitting galaxies by \cite{kennicutt_star_1998}:
\begin{equation}
    \mathrm{SFR} \ [\mathrm{M}_{\odot}/\mathrm{yr}] = 4.65 \times 10^{-42} \, \mathrm{L}(\mathrm{H}\alpha) \ [\mathrm{erg/s}]
\end{equation}
Note that we modify the relation for Chabrier IMF.

We derive gas-phase metallicities from the line flux ratio $\mathrm{N2} \equiv \log ([\mathrm{N\,{\footnotesize II}}]\lambda 6584/\mathrm{H}\alpha)$ using the calibration based on local galaxies \citep{pettini_o_2004}:
\begin{equation} \label{eq:N2}
    12 + \log (\mathrm{O/H}) = 8.90 + 0.57 \times \mathrm{N2}
\end{equation}
N2 calibration is the combination of strong emission lines adjacent to each other and therefore barely sensitive to dust attenuation and flux calibration.
The derived physical properties of H$\alpha$-detected sources are summarized in Table \ref{table:phys_value}.

Note that we classify AGN candidates based on multiple criteria.
First, we cross-match the samples against archival X-ray and radio sources in XCS2215 \citep{hilton_xmm_2010, klutse_meerkat_2024}.
\cite{klutse_meerkat_2024} investigate AGN populations selected with mid-infrared colour-colour criteria, far-infrared radio luminosity ratio, and far-infrared radio correlation for MeerKAT-detected galaxies in the cluster.
As a result, one source (ID:40) is confirmed to be an AGN-dominated galaxy detected by MeerKAT.
In addition, we use Baldwin--Phillips--Terlevich (BPT; \citeauthor{Baldwin_1981_classification_cn}\ \citeyear{Baldwin_1981_classification_cn}) diagram to separate AGN populations from star-forming galaxies for sources with available H$\beta$ and [O\,{\footnotesize III}]$\lambda$5007 data, obtained by archival MOIRCS spectroscopy \citep{hayashi_properties_2011} or our MOSFIRE observation (Figure\,\ref{fig:bpt}).
As a result, no objects are identified as AGNs although most of the sources are located near the boundary separating AGNs from star-forming galaxies.
Finally, for objects with either line ratios available, we define the sources satisfying at least one of the following criteria as AGN following \cite{kashino_fmos-cosmos_2017}: $\mathrm{FWHM}(\mathrm{H}\alpha) > 1000\mathrm{\, km/s}$,
$\mathrm{FWHM}([\mathrm{O\,{\footnotesize III}}]) > 1000\mathrm{\, km/s}$,
$\mathrm{N2} \geq -0.1$, $\mathrm{R3} \equiv \log ([\mathrm{O\,{\footnotesize III}}]\lambda 5007/\mathrm{H\beta}) \geq 0.9$.
As a result, we identify two AGN candidates (ID: 40, 42) in total in our sample.

\subsection{Spectral stacking} \label{sec:stacking}

\begin{figure*}[tbp]
    \begin{tabular}{cc}
      \begin{minipage}[t]{0.45\linewidth}
        \centering
        \includegraphics[width=1\linewidth]{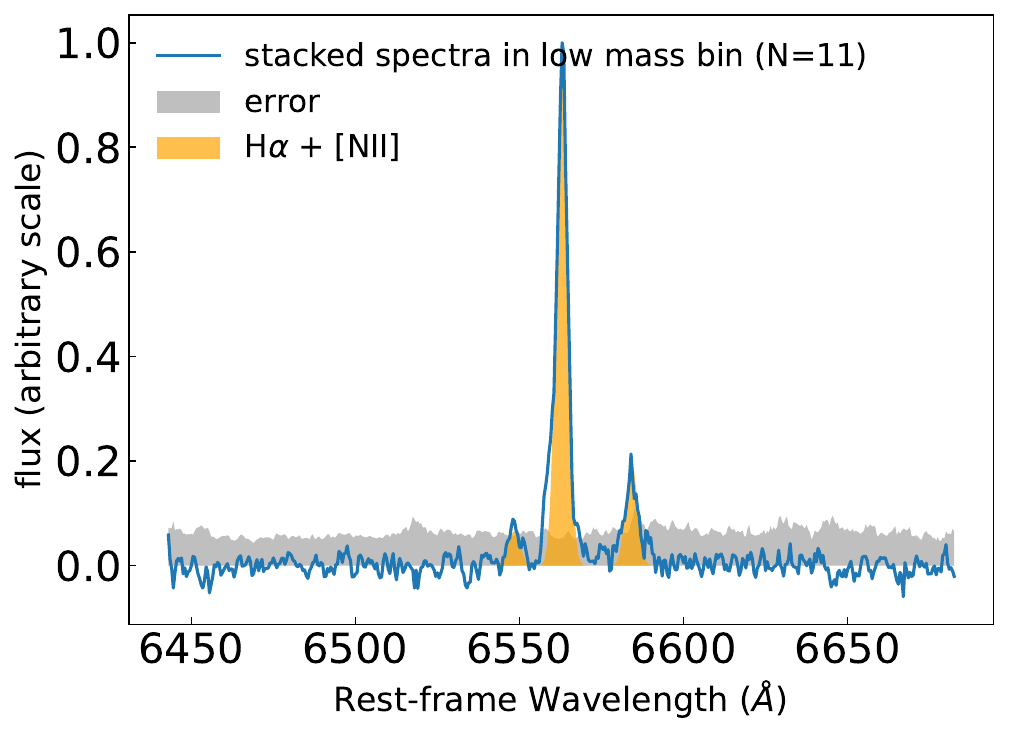}
      \end{minipage} &
      \begin{minipage}[t]{0.45\linewidth}
        \centering
        \includegraphics[width=1\linewidth]{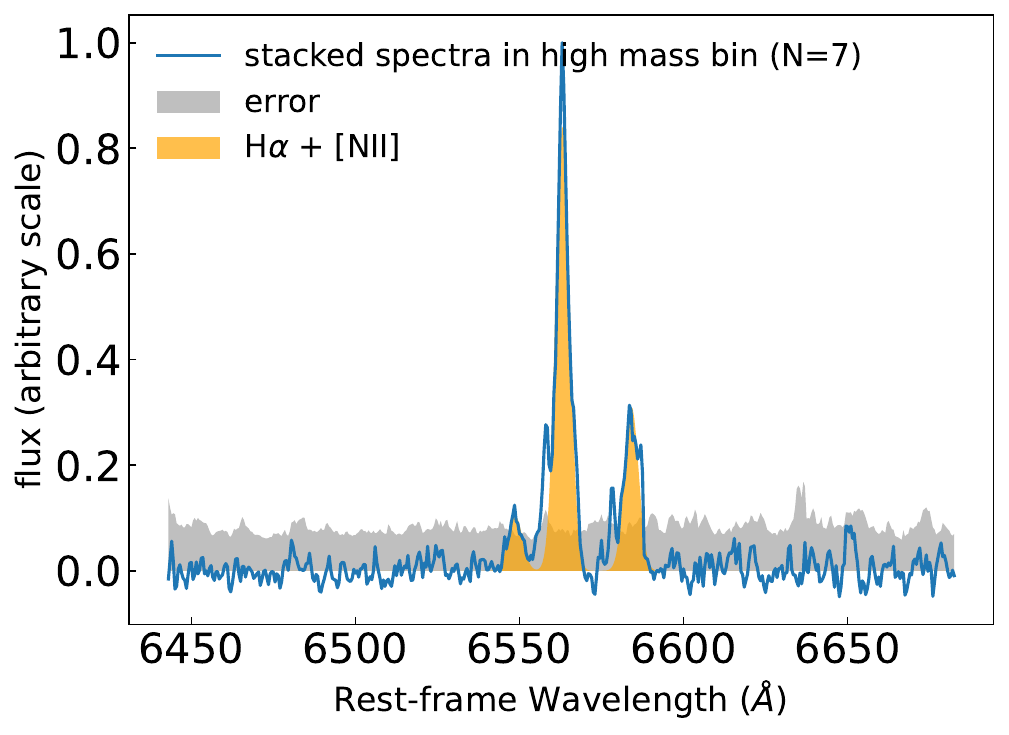}
      \end{minipage} \\

      \begin{minipage}[t]{0.45\linewidth}
        \centering
        \includegraphics[width=1\linewidth]{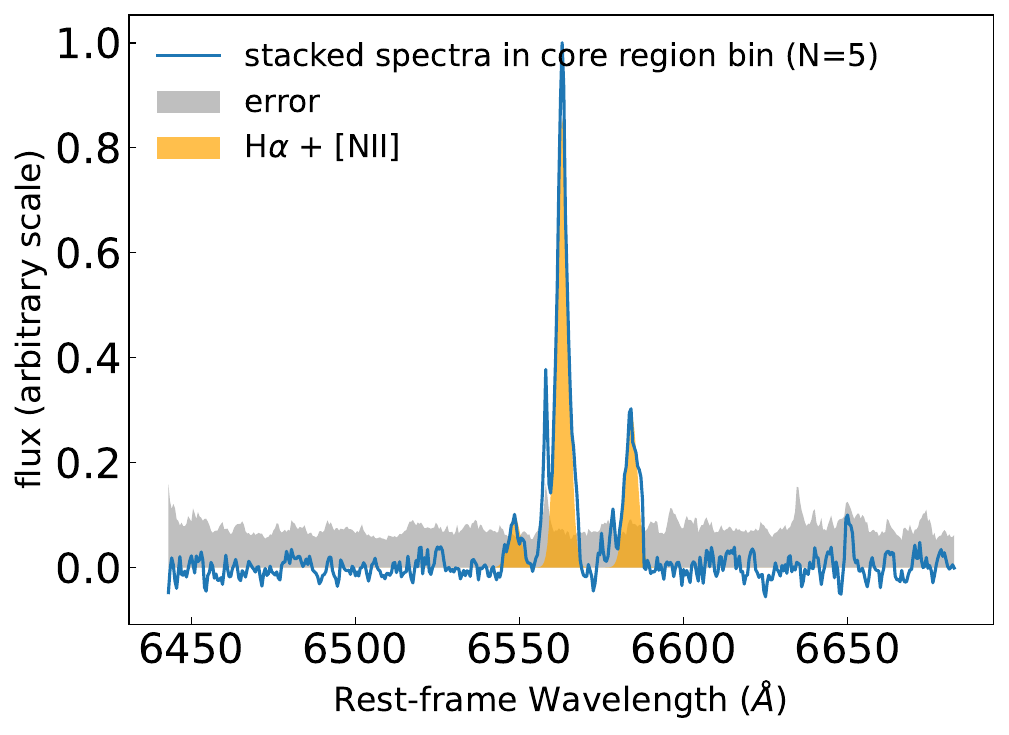}
        \label{fill}
      \end{minipage} &
      \begin{minipage}[t]{0.45\linewidth}
        \centering
        \includegraphics[width=1\linewidth]{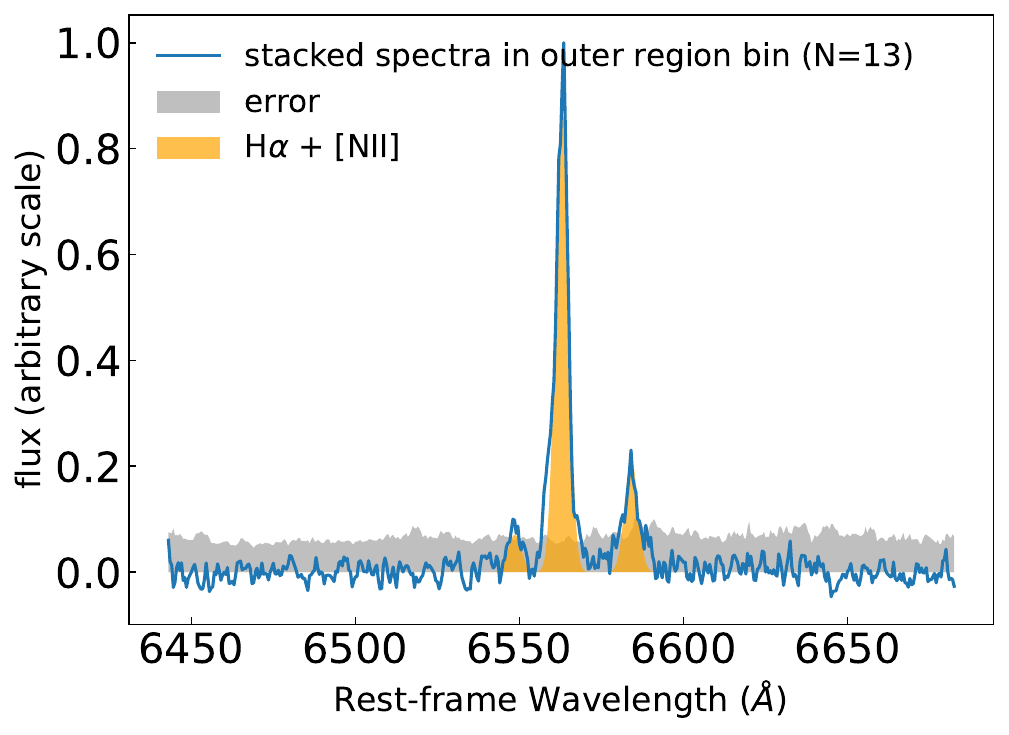}
      \end{minipage}

    \end{tabular}
    \caption{
        Stacked spectra (blue lines) around H$\alpha$+[N\,{\footnotesize II}]$\lambda\lambda 6548, 6583$  emission lines in the rest-frame wavelength for two stellar mass bins of $\log M_*/M_{\odot}<10.0$ (N=11; upper left) and  $\log M_*/M_{\odot} > 10.0$ (N=7; upper right), and two region bins of the core (N=5; lower left) and the outskirts (N=13; lower right).
        The gray region represents the stacked observational errors.
        Orange region show the fitted Gaussian function.
    }
    \label{fig:stacking}
\end{figure*}

\begin{table*}[tbp]
    \centering
    \caption{The physical properties derived from the stacked spectra of galaxies divided by stellar mass and region. $\Delta \mathrm{O/H}_\mathrm{cluster - field}$ represents the metallicity offset of each stacked value from field MZR \citep{kashino_fmos-cosmos_2017}.}
    \vspace{-5pt}
    \begin{tabular}{ccccccc}
        \hline
         &  N & $\log M_{*}/M_\odot$ range & Median $\log M_{*}/M_\odot$  &  $\log \mathrm{[N\,{\footnotesize II}]}\lambda 6584/\mathrm{H}\alpha$ & $12 + \log (\mathrm{O/H})$ & $\Delta \mathrm{O/H}_\mathrm{cluster - field}$ \\ \hline \hline
        Lower mass & 11 & $9.1 - 9.8$ &  9.6 & -0.67$\pm$0.07 & 8.52$\pm$0.04 & $0.14 \pm 0.06$\\ 
        Higher mass & 7 & $10.0 - 10.6$ &  10.3 & -0.43$\pm$0.04 & 8.65$\pm$0.02 & $0.08 \pm 0.05$ \\
        Core & 5 & $9.6 - 10.6$ &  10.1 & -0.46$\pm$0.05 & 8.64$\pm$0.02 & $0.09\pm0.05$\\
        Outskirts & 13 & $9.1 - 10.6$    &  9.7 & -0.62$\pm$0.06 & 8.55$\pm$0.03 & $0.15\pm0.06$\\ \hline
    \end{tabular}
    \label{tab:stacking}
\end{table*}

The metallicity analysis using N2 calibration is subject to a selection bias towards high metallicity sources for which [N\,{\footnotesize II}]$\lambda$6584 line flux is high enough to be detected.
To assess the environmental dependence of the metallicities for our samples without such a bias, we stack the observed spectra of H$\alpha$ detected sources including those with and without [N\,{\footnotesize II}]$\lambda$6584 emission line detections. 
Note that we exclude AGN candidates (Section \ref{sec:line-fitting}) from the analysis.
We divide the entire sample in two separate ways: by stellar mass, using stellar mass derived from SED fitting (Section \ref{sec:cigale}) and separated at $\log M_{*}/M_\odot = 10.0$, and by environment, distinguishing between the cluster core and outskirts as defined by phase-space diagram (Section\,\ref{sec:PS}).

We followed \cite{shimakawa_early_2015} for the stacking method.
The continuum-subtracted spectra of each target, with OH line remnants masked, are converted to rest-frame wavelengths, and aligned with the position of the H$\alpha$ emission line.
Then stacked spectra are derived as the following equation:
\begin{equation}
    F_{\mathrm{stack}} (\lambda) =
    \left. \sum_{i}^{n} \frac{F_i(\lambda)}{\sigma_i^2(\lambda)}
    \right/ \sum_{i}^{n} \frac{1}{\sigma_i^2(\lambda)}
\end{equation}
where i and n are the index and total number of stacked sources, and $F_i(\lambda)$ and $\sigma_i(\lambda)$ are the flux and its error for each target, respectively. 
The errors of the stacked spectra are derived by $\sigma_{\mathrm{stack}} (\lambda) = (\sum_i \frac{1}{\sigma(\lambda)_i^2})^{-\frac{1}{2}}$.
$3\sigma$ clipping for each wavelength is performed when stacking along with weighting by the observed error.
Then, line fitting and the metallicity derivation are conducted in the same manner as for individual sources.
The stacked spectra and the derived physical properties for each bin are shown in Figure \ref{fig:stacking} and Table \ref{tab:stacking}, respectively.

\subsection{SED fitting for estimation of stellar mass and dust attenuation} \label{sec:cigale}

To estimate stellar mass and dust extinction for our samples,
we conduct SED fitting for the photometric catalog constructed by \cite{hayashi_molecular_2018} (see Section \ref{sec:data_past}) using Code Investigating GALaxy Emission (CIGALE; \citeauthor{boquien_cigale_2019}\,\citeyear{boquien_cigale_2019}).
It is the Bayesian approach SED fitting code which takes into account the energy balance between UV-to-near-IR and mid- and far-IR.
For fitting, we use the photometry in the catalog including Subaru/Suprime-Cam \textit{B, Rc, i', z'} band and WFCAM $K$-band for all of our samples.
For a part of samples, we also use CFHT/MegaCam $u$ and $g$ bands, HST/WFC3 F125W, F140W, and F160W bands, CFHT/WIRCam $J$, $H$, $K$ bands, and Spitzer/IRAC [3.6], [4.5], and [5.8] channels, if available.
Photometry of these images was conducted by \cite{hayashi_molecular_2018} with a 2 arcsec diameter aperture and then applied the aperture correction of 0.43 mag, which is derived from the growth curve of PSF to be the total magnitude.

\begin{table*}[tbp]
    \centering
    \caption{SED fitting parameters that we use for the CIGALE code \citep{boquien_cigale_2019}.}
    \vspace{-5pt}
    \begin{tabular}{c|c}
        \hline
        Parameters & Values \\ \hline \hline
        Stellar population synthesis model & \cite{bruzual_stellar_2003} \\ 
        Initial mass function & \cite{chabrier_galactic_2003} \\ 
        Stellar metallicity & 0.02 \\ 
        Star formation history & exponentially declining with no burst: $\mathrm{SFR}(t) \propto t \exp(-t/\tau_\mathrm{SF})$ \\ 
        Star-forming time scale & $\log (\tau_\mathrm{SF}/\mathrm{[yr]}) = 8.5-10.0$ with steps of 0.1 \\ 
        Age of main stellar population & $\log (\tau_\mathrm{age}/\mathrm{[yr]}) = 8.0-9.7$ with steps of 0.1 \\ \hline
        Dust extinction law & \cite{calzetti_dust_2000} \\ 
        Dust attenuation & $A_V=0.0-3.0$ with steps of 0.3 \\ 
        Color excess ratio between stellar continuum and nebular emission & $E(B-V)_{\mathrm{stellar}} / E(B-V)_{\mathrm{nebular}} = 0.83$ \citep{kashino_fmos-cosmos_2013}\\ 
        Ratio of total to selective extinction & $R_V = 4.05$ \\ \hline
        additional error & 0.1 \\ \hline
    \end{tabular}
    \label{tab:cigale}
\end{table*}

The input parameters for the SED fitting are shown in Table \ref{tab:cigale}, and the estimated values are shown in Table \ref{table:phys_value}. 
The reduced $\chi^2$ ranges for $0.2 < \chi^2 < 4.2$ in our sample except for ID: 48 with reduced $\chi^2$ of $\sim28.6$.
In the following analyses, we ignore this galaxy and we use the derived stellar masses and dust extinctions for the remaining objects. 

\subsection{Phase-space diagram} \label{sec:PS}

\begin{figure}[tbp]
    \centering
    \includegraphics[width=1\linewidth]{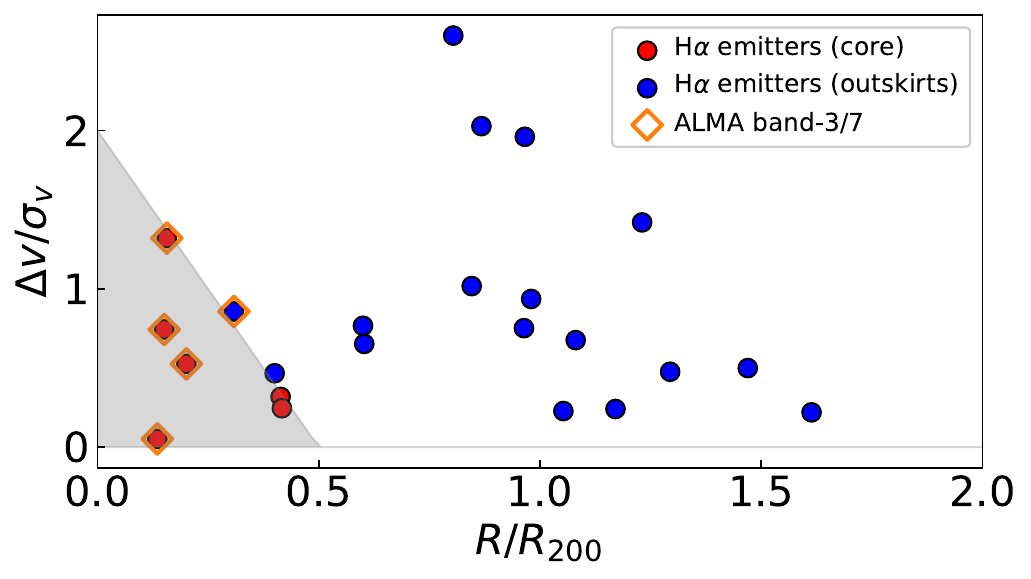}
    \caption{Phase-space diagram for H$\alpha$ emitters in XCS2215.
    We use the virial radius of $R_{200} = 1.23 \mathrm{\,Mpc}$ and velocity dispersion of $\sigma_v = 1128 \mathrm{\,km/s}$ \citep{maier_cluster_2019}.
    The gray area shows the virialized region defined by \cite{rhee_phase-space_2017}, regarded as the ``core" region in the cluster.
    Red and blue points represent the core and outskirts populations, respectively.
    Orange diamonds represent member galaxies observed by ALMA band-3/7.
    }\label{fig:PS}
\end{figure}

To analyze the accretion states of galaxies to the cluster,
we use a phase-space diagram (Figure \ref{fig:PS})
which enables us to avoid the projection effect and efficiently divide the samples into the core region and the outskirts.
Relative velocity $\Delta v = c|z_{\mathrm{obs}} - z_{\mathrm{cl}}| / (1+z_{\mathrm{cl}})$ and the projected distance $R$ from the X-ray center \citep{stanford_xmm_2006} are normalized to velocity dispersion $\sigma_v = 1128 \mathrm{\,km/s}$ and virial radius $R_{200} = 1.23 \mathrm{\,Mpc}$ of the cluster \citep{maier_cluster_2019}, respectively.
In this study, we define the virialized region \citep{rhee_phase-space_2017} as the ``core" and elsewhere as the ``outskirts" of the cluster.

\section{Result} \label{sec:result}

In this section, we present the star formation activity and chemical enrichment of the galaxies in XCS2215 cluster compared with field counterparts at the same redshift and also other clusters at higher redshifts ($z>2$).

\subsection{Star formation activities of member galaxies}

\begin{figure}[tbp]
    \centering
    \includegraphics[width=1\linewidth]{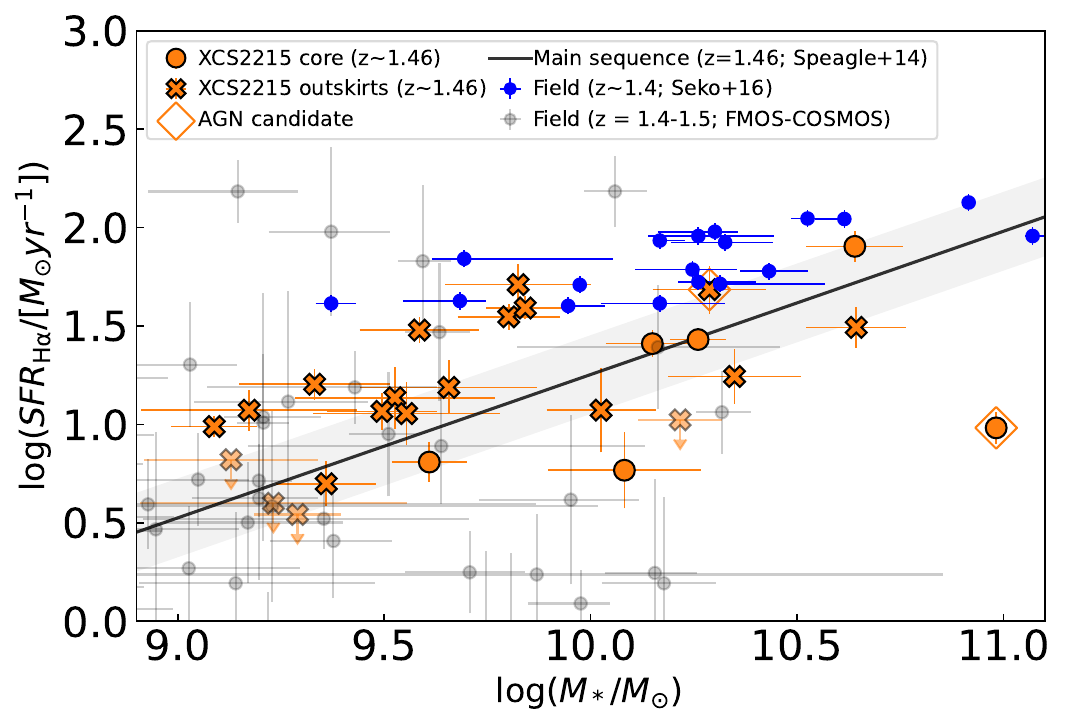}
    \caption{
        SFRs of galaxies in the XCS2215 cluster (orange points) derived from H$\alpha$ emission line fluxes.
        Here, we also show the galaxies without Ha detection ($\mathrm{S/N} <2$) as upper limits.
        The core and outskirts populations are shown as circles and crosses, respectively.
        AGN candidate are shown by points surrounded by diamond, which is classified by the method described in Section\,\ref{sec:line-fitting}.
        Gray points represent field galaxies in FMOS-COSMOS samples at $z\sim1.4-1.5$ \citep{silverman_fmos-cosmos_2015, kashino_fmos-cosmos_2019}.
        Blue points show the general field galaxies at $z\sim1.4$ \citep{seko_properties_2016}.
        The black line shows star-forming main sequence at $z\sim1.46$ \citep{speagle_highly_2014} with 0.2 dex scatter (gray region).
    } \label{fig:res/SFR_2215}
\end{figure}

Figure \ref{fig:res/SFR_2215} shows the dust-corrected SFR for each member galaxy derived from the method presented in Sections \ref{sec:line-fitting} and \ref{sec:cigale}.
For comparison, we plot the star-forming main sequence at $z\sim1.46$ (SFMS, \citeauthor{speagle_highly_2014}\,\citeyear{speagle_highly_2014}), FMOS-COSMOS samples at $z\sim1.4-1.5$ \citep{silverman_fmos-cosmos_2015, kashino_fmos-cosmos_2019}, and the sample of field galaxies at $z\sim1.4$ from \cite{seko_properties_2016}.
The SFMS \citep{speagle_highly_2014} is compiled from various relations in the literature using different SFR indicators.
In the latter two samples, SFRs are derived from $\mathrm{H\alpha}$ emission line fluxes.
For FMOS-COSMOS samples, we also correct dust attenuation by using Balmer decrement (H$\alpha$/H$\beta$) and adopt stellar mass provided in COSMOS2015 catalog \citep{Laigle_2016_cosmos2015_qz}, which is estimated using {\sc LePhare} code \citep{Arnouts_2002_measuring_jt, Ilbert_2006_accurate_sh}.

We find that SFRs of cluster galaxies lie around the SFMS for a given mass at the same redshift and are mostly found in the same regime as FMOS-COSMOS galaxies, especially at the low mass end.
We find no significant impact of their environment (core or outskirts) on star-forming activity for our samples.
We note that the samples of \cite{seko_properties_2016} are selected with $\mathrm{H\alpha}$ emission line fluxes larger than $1.0 \times 10^{16} \mathrm{erg \ s^{-1}cm^{-2}}$ to increase the success rate of $\mathrm{H\alpha}$ emission line detection \citep{yabe_nir_2012}, which results in the inherent selection bias toward higher SFR.
We also find that two H$\alpha$ emitters in the core lie below the SFMS.
One of them are particularly massive and an AGN candidate, whose star-forming activity is considered to be suppressed by AGN feedback.

We caveat that, given the fact that most of our samples are also selected as [O\,{\footnotesize II}] emitters, a bias towards higher SFR may still exist in our sample, too.
However, because all of the non-AGN member galaxies of XCS2215 for which [N\,{\footnotesize II}] is detected are above the lower limit of the $\mathrm{H\alpha}$ flux mentioned above, we can still make a fair comparison with \cite{seko_properties_2016} (see Section \ref{sec:FMR}, \ref{sec:outflow}).

\subsection{Chemical enrichment in XCS2215 cluster} \label{sec:mzr}

\begin{figure*}[tbp]
    \centering
    \includegraphics[width=1\linewidth]{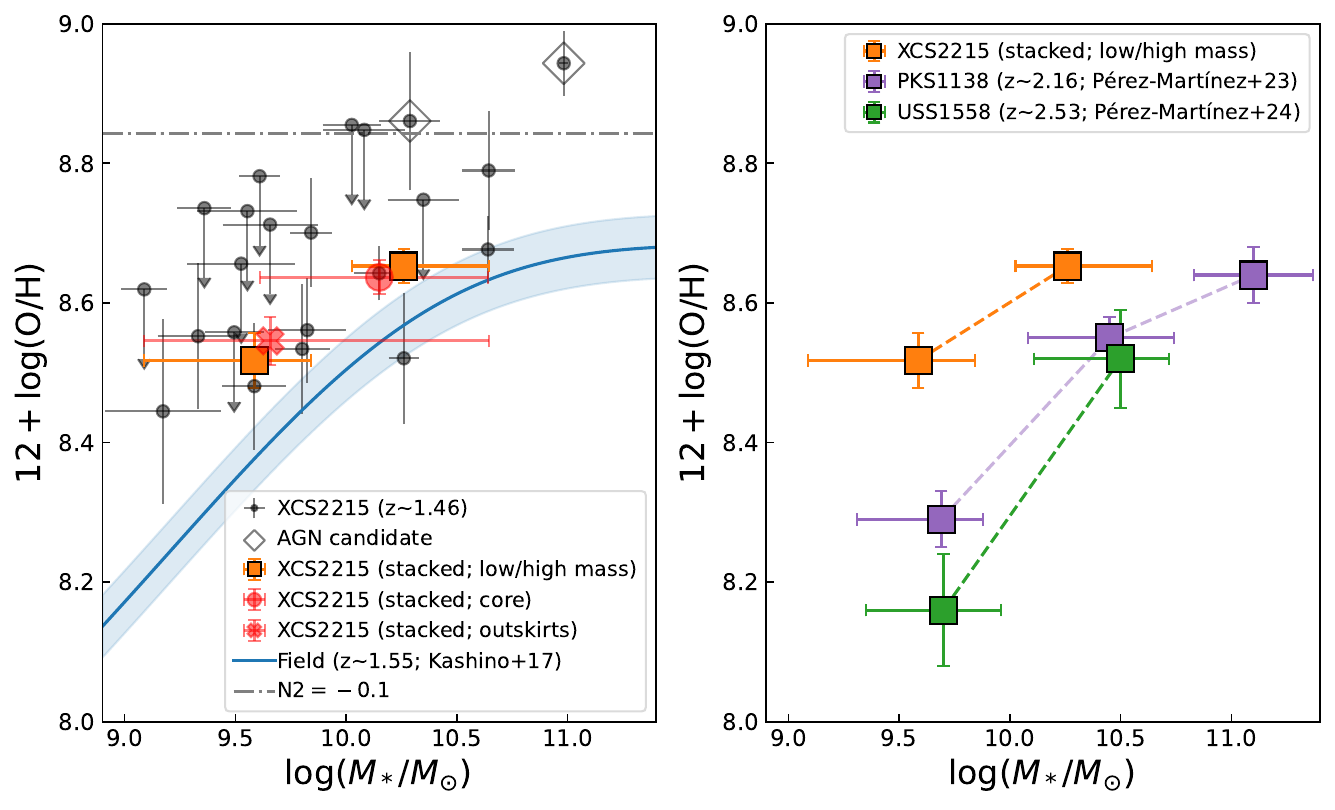}
    \caption{
        Left: Mass-metallicity relation (MZR) of XCS2215 cluster galaxies.
        Individual galaxies in the cluster are shown by black points.
        The blue line shows the metallicities of field galaxies at similar redshift derived by fitting to the stacked points of COSMOS field galaxies \citep{kashino_fmos-cosmos_2017}.
        Orange squares show the values derived from the stacked spectra of cluster galaxies divided into two mass bins, including [N\,{\footnotesize II}] non-detected objects (Sec \ref{sec:stacking}).
        Red point and cross show the stacked values of the core bin and outskirts bin, respectively.
        AGN candidates are shown as points surrounded by diamonds, which is classified by the method described in Section\,\ref{sec:line-fitting}, and the gray dash-dotted line represents the metallicity threshold to distinguish them \citep{kashino_fmos-cosmos_2017}.
        Right: The stacked gas-phase metallicities are compared to 2 protoclusters at $z>2$, namely, PKS1138 \citep{perez-martinez_signs_2023} and USS1558 \citep{perez-martinez_enhanced_2024}.
        The vertical error bar associated with stacked bins in both figures shows the stellar mass range of the galaxies within each bin.
    } \label{fig:MZR}
\end{figure*}

\begin{figure*}[tbp]
    \centering
    \includegraphics[width=1\linewidth]{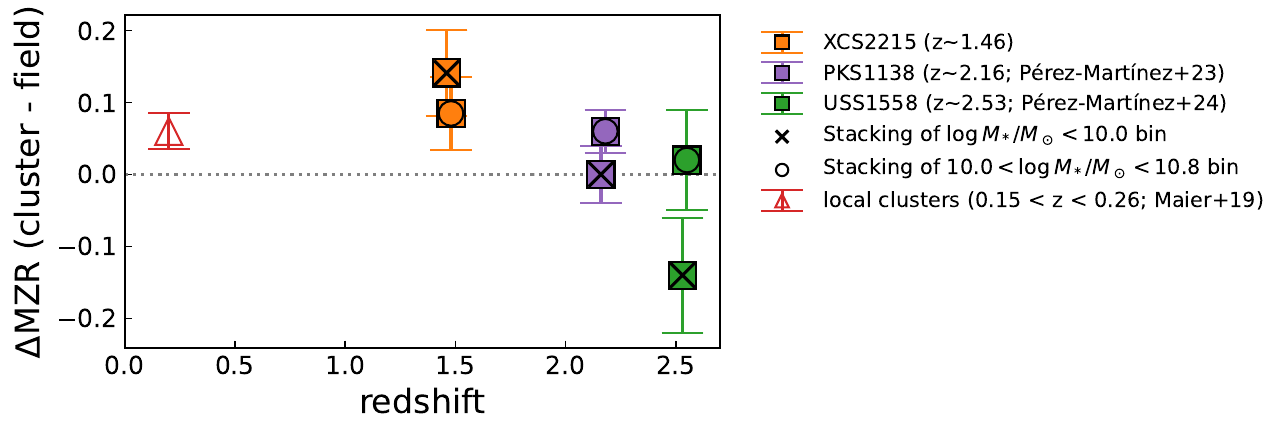}
    \caption{
        The redshift evolution of the metallicity offset for cluster galaxies with respect to the field MZR at each redshift.
        The color of points of $z \gtrsim 1.4$ protocluster corresponds to the stacked values in Figure \ref{fig:MZR}, and 
        the cross and circle symbols in each squares represent the stellar mass range of stacked bins of $\log M_*/M_\odot < 10.0 $ and $10.0 < \log M_*/M_{\odot} < 10.8$, respectively.
        The open red triangle shows the cluster galaxies at $0.15 < z < 0.26$ from LoCuSS survey \citep{maier_slow-then-rapid_2019}.
        The metallicity is derived from N2 calibration, except for the local value exploiting $\mathrm{O3N2} = \log \{[\mathrm{OIII}]\lambda 5007/\mathrm{H\beta})/([\mathrm{NII}]\lambda 6583/\mathrm{H\alpha})\}$ calibration.
    } \label{fig:dMZR-z}
\end{figure*}

Figure \ref{fig:MZR} shows the relation between gas-phase metallicities and stellar masses for individual galaxies in XCS2215 cluster and the values obtained from stacked spectra.
We compare our result with the MZR in the general field at a similar redshift by \cite{kashino_fmos-cosmos_2017}, in which they analyzed the MZR for 701 objects at $1.4 < z < 1.7$ with $\mathrm{H\alpha}$ detection in FMOS-COSMOS Survey \citep{silverman_fmos-cosmos_2015} using N2 calibration \citep{pettini_o_2004}.
This comparison indicates their metallicities are higher than those of the field galaxies at similar redshifts.
The offsets of the stacked values from the field relation are shown in Table \ref{tab:stacking}.
We note that the offset of the outskirts bin is larger than that of the core bin, contrary to the expectation that the core region, with high local density, might cause stronger environmental effects on the chemical evolution of member galaxies than in the outskirts.
For these bins, we investigate the cumulative distribution of stellar mass and local density,$\Sigma_5 = 5/ \pi D_4^2$, where $D_4$ is the distance to the 4th nearest neighbor confirmed members from each galaxy (Figure \ref{fig:CDF}).
As seen in the mass-divided bins in this study and in previous works \citep{kulas_mass-metallicity_2013, shimakawa_early_2015}, the metallicity offset from field galaxies tends to be larger at the lower mass end. 
This may be because low mass galaxies are more likely to lose metal content through gas stripping or outflows due to their shallower gravitational potential \citep{shimakawa_early_2015}.
Given that the core bin contains galaxies with higher stellar masses than those in the outskirts, the smaller metallicity offset in the core can be attributed to the difference in stellar mass distribution.
Additionally, the limited number of galaxies in each bin also contributes to this unexpected result.

The right panel of Figure \ref{fig:MZR} compares the results of this study with the stacked values of two higher redshift clusters, PKS1138 ($z\sim2.16$) and USS1558 ($z\sim2.53$) \citep{perez-martinez_signs_2023, perez-martinez_enhanced_2024}.
The metallicities in the referenced literature are all derived using N2 calibration.
The higher redshift clusters show lower metallicities, indicating the redshift evolution of the MZR.
Furthermore, the massive end shows much smaller evolution.
This indicates that the chemical evolution of galaxies with higher stellar masses proceeds earlier than that of galaxies with lower stellar masses (``down-sizing” scenario; \citeauthor{maiolino_amaze_2008} \citeyear{maiolino_amaze_2008}).
Figure \ref{fig:dMZR-z} shows the offset of stacked metallicities shown in the left panel of Figure \ref{fig:MZR} compared to the field galaxies at similar redshifts, plotted against redshift.
We see a clear redshift dependence for the lower mass bin at
$1.5 < z < 2.5$, where cluster galaxies show the positive metallicity offset from the field galaxies at $z$=1.5, while they show no or negative offset at $z>2$.
There is also a slight increasing trend for $0 < z < 1.5$, although the indices used for metallicity measurements are different between the lowest redshift bin (\citeauthor{maier_slow-then-rapid_2019}\,\citeyear{maier_slow-then-rapid_2019}; $\mathrm{O3N2} = \log \{[\mathrm{OIII}]\lambda 5007/\mathrm{H\beta})/([\mathrm{NII}]\lambda 6583/\mathrm{H\alpha})\}$) and all the others (N2), which may introduce a systematic difference.

\subsection{Environmental dependence evaluated by Fundamental Metallicity Relation} \label{sec:FMR}

\begin{figure*}[tbp]
    \centering
    \includegraphics[width=1\linewidth]{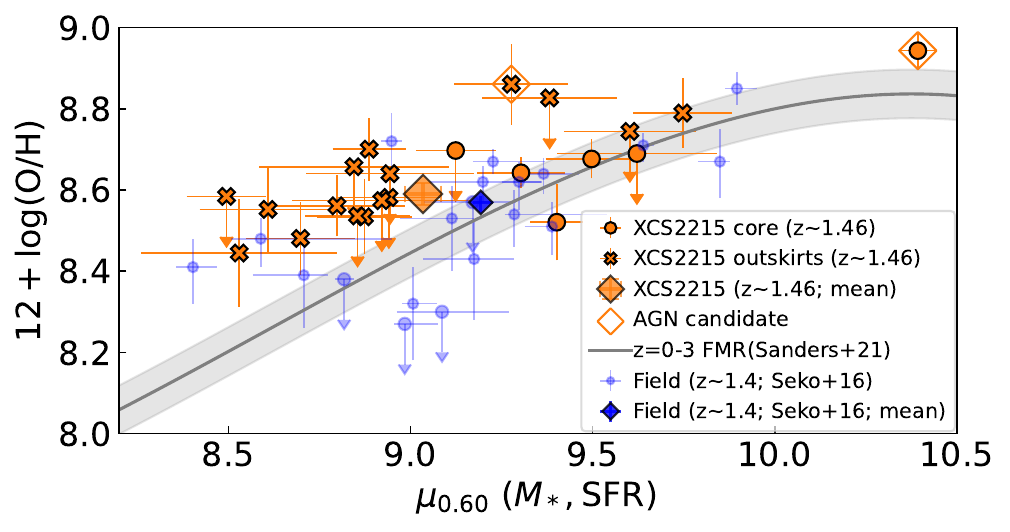}
    \caption{
    FMR for individual XCS2215 member galaxies. 
    The core and outskirts populations are shown as orange circles and crosses, respectively.
    The orange diamond shows their mean value.
    AGN candidate are shown by black points surrounded by diamond, which is classified by the method described in Section\,\ref{sec:line-fitting}.
    The black solid line represents the FMR for field galaxies at $0<z<3$ \citep{sanders_mosdef_2021}.
    Field galaxies at similar redshifts \citep{seko_properties_2016} and their mean value are also shown as blue points and diamond, respectively.
    }
    \label{fig:FMR_mu_060}
\end{figure*}

Next, we investigate the secondary dependence of gas-phase metallicity on SFR, which is the relation known as the fundamental metallicity relation (FMR; \citeauthor{mannucci_fundamental_2010}\,\citeyear{mannucci_fundamental_2010}; \citeauthor{Lara-Lopez_2010_fundamental_ai}\,\citeyear{Lara-Lopez_2010_fundamental_ai}).
Figure\,\ref{fig:FMR_mu_060} shows the gas-phase metallicity of the cluster galaxies as a function of $\mu_{\alpha} \equiv \log M_* - \alpha \times \log \mathrm{SFR}$, following \cite{mannucci_fundamental_2010} and \cite{sanders_mosdef_2021}.
Here $\alpha$ is defined as the value minimizing the scatter in metallicity-$\mu_{\alpha}$ space,
and we adopt $\alpha = 0.60$ from \cite{sanders_mosdef_2021}.

We compare our samples to the FMR for field galaxies from SDSS and MOSFIRE Deep Evolution Field (MOSDEF) survey ranging $0 < z < 3$ \citep{sanders_mosdef_2021}, which shows a tight relation with a significantly small scatter of 0.06\,dex. 
The difference between the mean value of the cluster galaxies excluding AGN candidates and the field FMR is significant with $\Delta(\mathrm{O/H}) = 0.13 \pm 0.07$.
From this result, we argue that the environmental dependence of metallicities seen in our samples (Section \ref{sec:mzr}) is not due to the primary contribution from more evolved and metal-enriched galaxies in the cluster compared to field environments, but is a result of the galaxy's surrounding environment itself.
We note that there is a slight offset in metallicities between the galaxies in the core and those in the outskirts, suggesting a potential environmental trend.
We caveat that the mean values are derived from the samples with [N\,{\footnotesize II}]$\lambda 6583$ detection.
Therefore, the selection bias toward higher metallicity is not completely eliminated.
However, the field galaxies from \cite{seko_properties_2016}, which are selected by the detection of H$\alpha$ emission line, the same way as for XCS2215, have values comparable to the global FMR \citep{sanders_mosdef_2021}.
Moreover, the individual data points of XCS2215 galaxies and the stack values are not very different in Figure\,\ref{fig:MZR}.
Therefore, we consider the selection bias to be not significant.

\cite{sanders_mosdef_2021} pointed out that there is a non-negligible effect depending on the selection of metallicity indices and calibration.
While this work and \cite{seko_properties_2016} use N2 index, 
FMR for \cite{sanders_mosdef_2021} is derived from metallicity using emission line ratios of [O\,{\footnotesize II}], H$\beta$, [O\,{\footnotesize III}], and [Ne\,{\footnotesize III}].
A larger sample size with a unified calibration method is needed to provide a more robust indication of the environmental dependence of chemical evolution.

\section{Discussion} \label{sec:disc}
\subsection{Gaseous outflow rate estimated by chemical evolution model} \label{sec:outflow}

\begin{figure*}[tbp]
    \centering
    \includegraphics[width=1\linewidth]{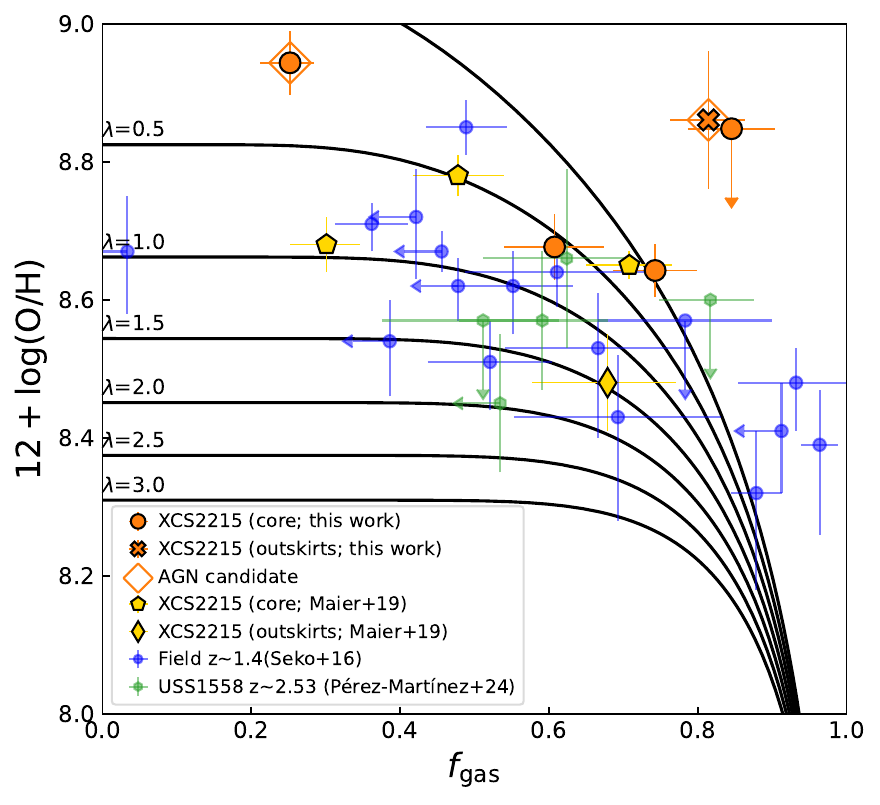}
    \caption{
        The gas-phase metallicity as a function of gas mass fraction for XCS2215 cluster galaxies.
        The core and outskirts populations are shown as orange circles and crosses, respectively.
        AGN candidate are shown by black points surrounded by diamond, which is classified by the method described in Section\,\ref{sec:line-fitting}.
        Yellow pentagons and diamonds show additional cluster members from \cite{maier_cluster_2019}, located in the core and outskirts, respectively.
        As a comparison, field galaxies at $z\sim1.4$ \citep{seko_properties_2016} and protocluster galaxies in USS1558 at $z\sim 2.53$ \citep{perez-martinez_enhanced_2024} are shown as blue and green points, respectively.
        The black lines represent the tracks of the gas regulator models for various mass-loading factor $\lambda$ ranging from 0.0 to 3.0 with a step of 0.5 (from top to bottom).
        }\label{fig:z_fgas}
\end{figure*}

In order to investigate the physical mechanisms to cause the environmental dependence of chemical enrichment that we see,
we utilize the gas regulator model \citep{lilly_gas_2013, peng_haloes_2014}, which takes into account the chemical evolution incorporating star formation, metal production, gaseous inflow, and outflow.
In this model, the normalization of gas-phase metallicities for a given gas mass fraction $f_\mathrm{gas} = M_\mathrm{gas} / (M_\mathrm{gas} + M_\mathrm{stellar})$ depends mainly on the mass-loading factor $\eta \equiv \mathrm{outflow \ rate}/\mathrm{star \ formation \ rate}$ (see figure \ref{fig:z_fgas}), which enables us to constrain the outflow rate of the system. 

We follow \cite{suzuki_dust_2021} for the assumption of parameters and initial conditions of the model.
We assume the return mass fraction $R$ = 0.4 in Charbrier IMF \citep{madau_cosmic_2014} and the metallicity yield $y = 1.5 Z_{\odot}$.
Gaseous inflow into the system is assumed to be metal free ($Z_0=0$). 
We set star formation efficiency $\epsilon$ and the gas consumption timescale $\tau_{\mathrm{dep}}$ to be $\tau_{\mathrm{dep}}= 1/\epsilon = 0.8 \mathrm{Gyr}$.
It should be noted that the normalization of the model tracks does not significantly depend on $\tau_{\mathrm{dep}}$ \citep{suzuki_dust_2021}.

Figure \ref{fig:z_fgas} illustrates the gas regulator model with values of the mass-loading factor $\eta$ ranging from 0.0 to 3.0 with a step of 0.5 (from top to bottom) for five cluster members for which ALMA gas measurements are available and H$\alpha$ and (+[N\,{\footnotesize II}]) are detected.
Those five galaxies are shown with measured gas mass fraction and metallicities (one of them is upper limit and two of them are AGN candidates). 
We also plot that six additional cluster members with ALMA gas measurements, for which the gas-phase metallicities are obtained using VLT/KMOS $H$-band spectroscopy \citep{maier_cluster_2019}.
For comparison, field galaxies at similar redshifts \citep{seko_properties_2016}, and member galaxies in USS1558 cluster at $z\sim2.53$ \citep{perez-martinez_enhanced_2024} are also shown.
While one cluster member in the outskirts, field galaxies, and the USS1558 galaxies are spread in the range of $0.0 < \eta < 2.0$, the five galaxies located in the central virialized region of the cluster have smaller values, $0.0 < \eta < 1.0$.
The outflows of these sources are likely to be weaker than those of the field counterparts, suggesting that the unique environment of the galaxy cluster invokes such an environmental dependence.
However, it should be noted that the sample may be biased toward high metallicity and high gas fraction because here we exclude the galaxies for which [N\,{\footnotesize II}] or CO emission line is not detected.

\subsection{Gas-phase metallicities enhancement in cluster environment}

The MZR (Figure \ref{fig:MZR}) shows that the metallicity in ISM in the cluster galaxies at $z\sim1.5$ are systematically higher than those in field galaxies at similar redshifts for a given mass.
Enhanced metallicity in galaxies residing in clusters or overdensities have been reported both in observational studies from local to high redshifts \citep{shimakawa_early_2015, maier_clash-vlt_2016,wu_dependence_2017, maier_cluster_2019,ciocan_clash-vlt_2020,donnan_role_2022, perez-martinez_signs_2023} and in simulation studies \citep{wang_environmental_2023, oku_osaka_2024}.
Such an enhancement of metallicity is thought to be the result of the following effects, and the combinations thereof:
(i) During infalling to the cluster center, the ram pressure by ICM and tidal forces due to the galaxy-galaxy interaction can strip the low-metallicity gas that is weakly bounded in the outer region of the disk \citep{maiolino_re_2019}, increasing the overall gas-phase metallicity integrated across the galaxies \citep{shimakawa_early_2015, khoram_gas_2024}.
(ii) Low-density and low-metallicity neutral gas in the halo reservoir is stripped away when falling in, and the accretion of the pristine gas onto the disk is suppressed as a result (``strangulation"; \citeauthor{larson_evolution_1980}\ \citeyear{larson_evolution_1980}; \citeauthor{balogh_origin_2000}\ \citeyear{balogh_origin_2000}; \citeauthor{alberts_clusters_2022}\ \citeyear{alberts_clusters_2022}).
The galaxy gently consumes the remaining gas in the disk for star formation, and chemical evolution proceeds without dilution, increasing the entire metallicity.
(iii) When chemically enriched gas is ejected out of the ISM by the outflow, it may be pushed back into the ISM by the pressure of the surrounding ICM.
It has been proposed that the metallicity of the ISM can be further enhanced by recycling the re-accreted, already chemically enriched gas, leading to a smaller mass-loading factor \citep{dave_galaxy_2011,kulas_mass-metallicity_2013,shimakawa_early_2015, perez-martinez_signs_2023, perez-martinez_enhanced_2024}. 

The cluster galaxies have a relatively high SFR and molecular gas fraction compared to field galaxies \citep{seko_properties_2016} based on Figure\,\ref{fig:res/SFR_2215} and \ref{fig:z_fgas}.
Therefore, these results in XCS2215 cluster galaxies favor the strangulation scenario rather than the gas stripping scenario in the disk.
The previous study of XCS2215 using VLT/KMOS spectroscopy, \cite{maier_cluster_2019}, also supports the strangulation scenario based on the high gas fraction and SFR to interpret the metallicity enhancement for the galaxies inside $0.5 R_{200}$.
Our result is also consistent with the outflow confinement scenario due to the elevated pressure of hot ICM (see Section\,\ref{sec:high-z_clusters}) as the member galaxies residing in the cluster core have a lower mass-loading factor of $0.0 < \lambda < 0.5$ than field galaxies (Section\,\ref{sec:outflow}).
We still need further observations to determine whether strangulation or outflow confinement is primarily responsible for the chemical enhancement of these galaxies.

\subsection{Comparison to other cluster galaxies at higher redshift} \label{sec:high-z_clusters}

Figure \ref{fig:MZR} reveals that the galaxies in XCS2215 at $z\sim1.46$ have higher metallicity values than those of the cluster galaxies in PKS1138 \citep{perez-martinez_signs_2023} and USS1558 \citep{perez-martinez_enhanced_2024} at higher redshift ($z\gtrsim2$).
On top of that, Figure \ref{fig:dMZR-z} shows the metallicity offset for these protocluster galaxies compared to the field counterparts at each redshift, which reveals that the offset evolves with redshift, especially in the lower stellar mass range ($\log M_*/M_\odot \leq 10.0$). 
This may reflect the difference in the evolutionary stages among these protoclusters at different redshifts.
In protoclusters at $z\gtrsim2$, cold streams from surrounding large-scale structures can penetrate through the ICM to galaxies, efficiently feeding star formation activities in the galaxies.
On the other hand, in clusters/protoclusters at $z<2$, the accretion of cold gas becomes inefficient by shock-heating of the ICM in deeper potential wells,
and the pristine gas supply to the cluster galaxies is eventually depleted \citep{dekel_galaxy_2006,overzier_realm_2016,shimakawa_mahalo_2018-1}.
In this context, XCS2215 appears to be a relatively mature cluster, and the cold gas accretion to the galaxies is probably suppressed.
These processes may result in less dilution of the gas-phase metallicity in ISM compared to the $z\gtrsim2$ clusters.
This picture is consistent with XCS2215 being an X-ray detected cluster, and that the environmental dependence of the MZR is prominent in XCS2215 (Figure \ref{fig:dMZR-z}).
Additionally, the virial mass $M_{200}$ of XCS2215 ($M_{200} \sim 6.3 \times 10^{14} M_{\odot}$; \citeauthor{maier_cluster_2019}\ \citeyear{maier_cluster_2019}) shows a higher value than that of PKS1138 ($M_{200} \sim 1.7 \times 10^{14} M_{\odot}$) and USS1558 ($M_{200} < 8.7 \times 10^{13} M_{\odot}$), even when their mass growth with redshift is taken into account \citep{shimakawa_identification_2014}.
This comparison also supports that the mass assembly drives the transition of gas accretion phase from evolving protoclusters at $z\gtrsim2$ to mature clusters at lower redshift, and XCS2215 is the later stage of the evolutionary phases.

\section{Conclusion} \label{sec:concl}

We conduct NIR spectroscopic observations of the galaxies in X-ray detected cluster XCS2215 at $z\sim1.46$ using the Keck/MOSFIRE and analyzed 23 objects for which H$\alpha$ emission line is detected.
In combination with the existing ALMA data and a simple analytical model, we also investigate the chemical evolution in these galaxies and their environmental dependence at cosmic noon.

The stacked metallicity of the member galaxies is found to be 0.08--0.14 dex higher compared to field galaxies at similar redshifts.
In comparison with PKS1138 ($z\sim2.16$) and USS1558 ($z\sim2.53$), gaseous metallicities in XCS2215 ($z\sim1.46$) are on average higher than those of the higher redshift clusters ($z\sim2$). These results indicate a stronger environmental impact on chemical evolution in the XCS2215 cluster.

The metallicity enhancement in XCS2215 can be attributed to the strangulation, the stripping of gas in the halos by the ICM or interactions.
In addition, in the central part of the cluster, the outflow is confined by the ICM pressure which pushes back the outflowing gas to the disks, resulting in recycling of the enriched gas for the next episode of star formation and thus accelerating chemical evolution in these galaxies. 
Lastly, compared to other cluster galaxies at higher redshifts, in terms of the growth stage of the cluster, XCS2215 is thought to be a relatively mature cluster with the advanced evolutionary stage.
Therefore, the inflow of pristine gas into the cluster halo from surrounding large-scale structures is already inefficient with the ICM in a hot mode as a X-ray cluster, leading to the lack of metallicity dilution.

It should be noted, however, that the investigation of the environmental dependence at high redshifts is yet statistically insufficient, and further work is needed to confirm our results. 
The upcoming telescopes and instruments such as Subaru/PFS, VLT/MOONS and ELTs will efficiently provide us with a much larger spectroscopic sample of protocluster galaxies in various evolutionary stages leveraging their high sensitivities and/or wide field coverages.

\section*{Acknowledgments}
This work was supported by JSPS Core-to-Core Program (grant number: JPJSCCA20210003).
TK acknowledges financial support from JSPS KAKENHI Grant Numbers 24H00002 (Specially Promoted Research by T. Kodama et al.) and 22K21349 (International Leading Research by S. Miyazaki et al.).
JMPM acknowledges funding from the European Union's Horizon-Europe research and innovation program under the Marie Sk{\l}odowska-Curie grant agreement No. 101106626.
MO acknowledges financial support from JSPS KAKENHI Grant Numbers 25K07361.
Some of the data presented herein were obtained at Keck Observatory, which is a private 501(c)3 non-profit organization operated as a scientific partnership among the California Institute of Technology, the University of California, and the National Aeronautics and Space Administration. The Observatory was made possible by the generous financial support of the W. M. Keck Foundation.
The authors wish to recognize and acknowledge the very significant cultural role and reverence that the summit of Maunakea has always had within the Native Hawaiian community. We are most fortunate to have the opportunity to conduct observations from this mountain.
The observations were carried out within the framework of Subaru-Keck/Subaru-Gemini time exchange program which is operated by the National Astronomical Observatory of Japan. We are honored and grateful for the opportunity of observing the Universe from Maunakea, which has the cultural, historical and natural significance in Hawaii.
This paper makes use of the following ALMA data: ADS/JAO.ALMA\#2011.1.00623.S and ADS/JAO.ALMA\#2015.1.00779.S.
ALMA is a partnership of ESO (representing its member states), NSF (USA) and NINS (Japan), together with NRC (Canada), NSTC and ASIAA (Taiwan), and KASI (Republic of Korea), in cooperation with the Republic of Chile.
The Joint ALMA Observatory is operated by ESO, AUI/NRAO and NAOJ.
This research is based in part on data collected at the Subaru Telescope, which is operated by the National Astronomical Observatory of Japan. We are honored and grateful for the opportunity of observing the Universe from Maunakea, which has the cultural, historical, and natural significance in Hawaii.
Based on observations obtained with MegaPrime/MegaCam, a joint project of CFHT and CEA/IRFU, at the Canada-France-Hawaii Telescope (CFHT) which is operated by the National Research Council (NRC) of Canada, the Institut National des Science de l'Univers of the Centre National de la Recherche Scientifique (CNRS) of France, and the University of Hawaii. This work is based in part on data products produced at Terapix available at the Canadian Astronomy Data Centre as part of the Canada-France-Hawaii Telescope Legacy Survey, a collaborative project of NRC and CNRS.
Some of the data presented in this paper were obtained from the Multimission Archive at the Space Telescope Science Institute (MAST). STScI is operated by the Association of Universities for Research in Astronomy, Inc., under NASA contract NAS5-26555. Support for MAST for non-HST data is provided by the NASA Office of Space Science via grant NAG5-7584 and by other grants and contracts.
Based on data products from observations made with ESO Telescopes at the La Silla Paranal Observatory under ESO programme ID 179.A-2005 and on data products produced by TERAPIX and the Cambridge Astronomy Survey Unit on behalf of the UltraVISTA consortium.
This work made use of Astropy:\footnote{\url{http://www.astropy.org}} a community-developed core Python package and an ecosystem of tools and resources for astronomy \citep{astropy:2013, astropy:2018, astropy:2022}.

\facilities{Keck:I(MOSFIRE), ALMA, CFHT, Subaru, HST, Spitzer, UKIRT}

\software{astropy \citep{astropy:2013, astropy:2018, astropy:2022}, Numpy \citep{Harris2020-ra}, SciPy \citep{Virtanen2020-zv}}

\appendix
\section{Physical properties} \restartappendixnumbering

\begin{table*}[htbp]
    \vspace{3.5cm}
    \begin{rotatetable*}
        \centering
        \caption{Physical properties of H$\alpha$-detected member galaxies in XCS2215 used in this work.}
        \begin{tabular}{rllcrccccccc}
            \hline
            ID & R.A.(J2000) & Dec.(J2000) & z\tablenotemark{\small{\rm a}} & $\log M_{*}/M_{\odot}$\tablenotemark{\small{\rm b}} & $E(B-V)_{\mathrm{stellar}}$\tablenotemark{\small{\rm b}} & $\log \mathrm{SFR}[\mathrm{M}_{\odot} \ yr^{-1}]$\tablenotemark{\small{\rm a}} & N2\tablenotemark{\small{\rm c}} & R3\tablenotemark{\small{\rm d}} & $12+\log \mathrm{(O/H)}$\tablenotemark{\small{\rm e}} &  $\log M_\mathrm{mol}/M_{\odot}$\tablenotemark{\small{\rm f}} & $f_\mathrm{gas}$ \\ \hline
2&$333.9562426$&$-17.6494988$&$1.4639$&$9.36\pm0.12$&$0.14\pm0.03$&$0.70\pm0.11$&$<-0.29$&$>0.50^\mathrm{*}$&$<8.74$&&\\
3&$333.9576805$&$-17.6536936$&$1.4657$&$9.09\pm0.10$&$0.14\pm0.02$&$0.99\pm0.05$&$<-0.49$&$>-1.42$&$<8.62$&&\\
4&$333.9580179$&$-17.6737249$&$1.4614$&$9.53\pm0.24$&$0.34\pm0.06$&$1.14\pm0.16$&$<-0.43$&&$<8.66$&&\\
5&$333.9598156$&$-17.6310796$&$1.4810$&$9.49\pm0.13$&$0.24\pm0.04$&$1.07\pm0.09$&$<-0.60$&&$<8.56$&&\\
7&$333.9686310$&$-17.6294998$&$1.4630$&$10.64\pm0.12$&$0.40\pm0.05$&$1.49\pm0.10$&$-0.19\pm0.15$&&$8.79\pm0.09$&&\\
11&$333.9750274$&$-17.6504455$&$1.4641$&$10.35\pm0.16$&$0.37\pm0.05$&$1.24\pm0.14$&$<-0.27$&&$<8.75$&&\\
12&$333.9782501$&$-17.6418735$&$1.4599$&$9.61\pm0.09$&$0.15\pm0.03$&$0.81\pm0.10$&$<-0.21$&&$<8.78$&&\\
13&$333.9788422$&$-17.6703626$&$1.4751$&$9.82\pm0.18$&$0.41\pm0.04$&$1.71\pm0.10$&$-0.59\pm0.13$&&$8.56\pm0.08$&&\\
16&$333.9808485$&$-17.6659104$&$1.4664$&$9.80\pm0.12$&$0.28\pm0.04$&$1.55\pm0.06$&$-0.64\pm0.16$&&$8.53\pm0.09$&&\\
21&$333.9868618$&$-17.5995539$&$1.4383$&$9.33\pm0.18$&$0.22\pm0.04$&$1.20\pm0.08$&$-0.61\pm0.18$&$0.45\pm0.15$&$8.55\pm0.10$&&\\
22&$333.9871323$&$-17.6346675$&$1.4448$&$10.15\pm0.11$&$0.27\pm0.04$&$1.41\pm0.07$&$-0.45\pm0.07$&&$8.64\pm0.04$&$10.61^{0.06}_{0.06}$&$0.74^{0.06}_{0.06}$\\
25&$333.9891347$&$-17.6177684$&$1.4593$&$10.26\pm0.07$&$0.29\pm0.02$&$1.43\pm0.04$&$-0.67\pm0.17$&$0.19\pm0.16$&$8.52\pm0.09$&&\\
31&$333.9926633$&$-17.6393985$&$1.4575$&$10.64\pm0.12$&$0.55\pm0.04$&$1.90\pm0.08$&$-0.39\pm0.08$&&$8.68\pm0.05$&$10.83^{0.03}_{0.04}$&$0.61^{0.07}_{0.07}$\\
33&$333.9947404$&$-17.6280017$&$1.4639$&$10.08\pm0.19$&$0.31\pm0.06$&$0.77\pm0.19$&$<-0.09$&&$<8.85$&$10.82^{0.05}_{0.06}$&$0.85^{0.06}_{0.06}$\\
34&$333.9951625$&$-17.6776769$&$1.4632$&$9.17\pm0.26$&$0.16\pm0.06$&$1.07\pm0.10$&$-0.80\pm0.23$&$0.49\pm0.15$&$8.44\pm0.13$&&\\
35&$333.9956374$&$-17.6837220$&$1.4701$&$10.03\pm0.13$&$0.47\pm0.04$&$1.07\pm0.21$&$<-0.08$&&$<8.86$&&\\
40\tablenotemark{\small{\rm g}}&$334.0015457$&$-17.6307469$&$1.4521$&$10.98\pm0.02$&$0.08\pm0.03$&$0.98\pm0.08$&$0.08\pm0.08$&&$8.94\pm0.05$&$10.51^{0.07}_{0.09}$&$0.25^{0.03}_{0.04}$\\
42\tablenotemark{\small{\rm g}}&$334.0038368$&$-17.6419642$&$1.4649$&$10.29\pm0.14$&$0.56\pm0.05$&$1.69\pm0.12$&$-0.07\pm0.17$&&$8.86\pm0.10$&$10.93^{0.03}_{0.05}$&$0.81^{0.05}_{0.05}$\\
43&$334.0106336$&$-17.6323017$&$1.4613$&$9.84\pm0.09$&$0.44\pm0.02$&$1.59\pm0.06$&$-0.35\pm0.14$&&$8.70\pm0.08$&&\\
47&$334.0375368$&$-17.6423564$&$1.4591$&$9.59\pm0.14$&$0.27\pm0.04$&$1.48\pm0.07$&$-0.74\pm0.16$&$0.45\pm0.45^\mathrm{*}$&$8.48\pm0.09$&&\\
48\tablenotemark{\small{\rm h}}&$334.0422327$&$-17.6238621$&$1.4592$&&&&$<-0.22$&&$<8.78$&&\\
49&$334.0560355$&$-17.6359197$&$1.4616$&$9.66\pm0.21$&$0.41\pm0.05$&$1.19\pm0.14$&$<-0.33$&&$<8.71$&&\\
50&$334.0611271$&$-17.6226245$&$1.4590$&$9.55\pm0.23$&$0.30\pm0.06$&$1.06\pm0.16$&$<-0.30$&&$<8.73$&&\\
        \hline
        \end{tabular}
        \noindent\parbox{\textwidth}{
            \textbf{Notes.} \\
            $^{\rm a}$ Spectroscopic redshift measured by H$\alpha$ emission lines. \\
            $^{\rm b}$ Derived from SED fitting (Section\ \ref{sec:cigale}). \\
            $^{\rm c}$ $\mathrm{N2} = \log ([\mathrm{N\,{\footnotesize II}}]\lambda 6584/\mathrm{H}\alpha)$. \\
            $^{\rm d}$ $\mathrm{R3} = \log ([\mathrm{O\,{\footnotesize III}}]\lambda 6584/\mathrm{H}\beta$) derived by either MOSFIRE/J-band observation in this work or MOIRCS/J-band observation in \cite{hayashi_properties_2011}. The latter cases are indicated by asterisk (*). \\
            $^{\rm e}$ Based on N2 calibration. \\
            $^{\rm f}$ Derived from CO(J=2-1) emission lines measured by ALMA band-3 observation \citep{hayashi_evolutionary_2017}. \\
            $^{\rm g}$ AGN candidates excluded from the stacking analysis (Section~\ref{sec:line-fitting}). \\
            $^{\rm h}$ Excluded from the analysis due to large reduced $\chi^2$ ($\sim28.6$; Section\ \ref{sec:cigale}).
        }
    \label{table:phys_value}
    \end{rotatetable*}
\end{table*}

\clearpage
\section{Line fitting result} \restartappendixnumbering \label{sec:app-line} 

\begin{figure*}[h]
    \centering
    \begin{minipage}{0.22\textwidth}\centering\includegraphics[width=\linewidth]{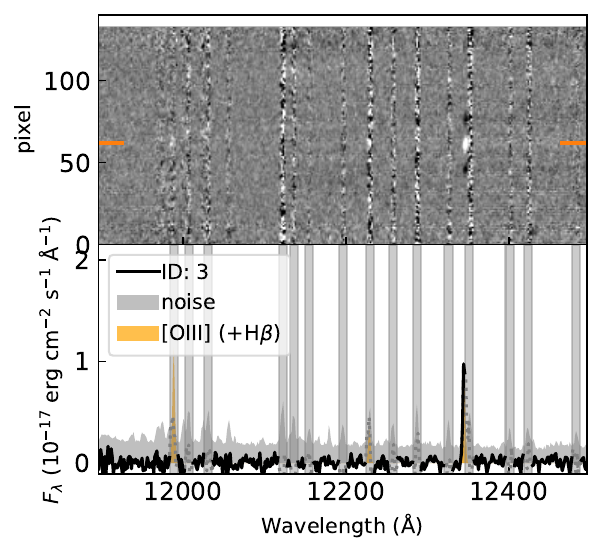}\end{minipage}\hfill
    \begin{minipage}{0.22\textwidth}\centering\includegraphics[width=\linewidth]{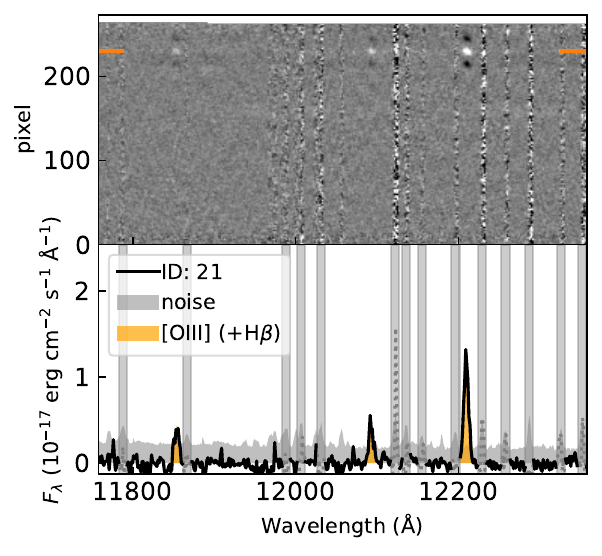}\end{minipage}\hfill
    \begin{minipage}{0.22\textwidth}\centering\includegraphics[width=\linewidth]{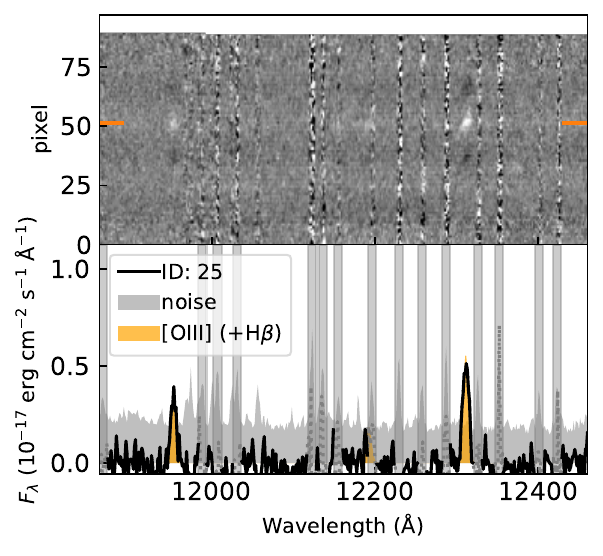}\end{minipage}\hfill
    \begin{minipage}{0.22\textwidth}\centering\includegraphics[width=\linewidth]{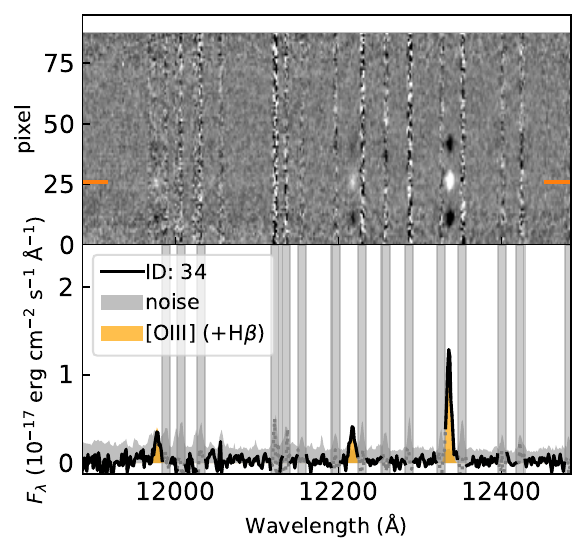}\end{minipage}
    \caption{Top: 2D \textit{J}-band spectra of H$\alpha$-detected galaxies in XCS2215 after the reduction using MOSFIRE DRP (Section\,\ref{observation}), showing the detected [O\,{\footnotesize III}]$\lambda$5007 emission lines.
    The 1D extraction is performed using the pixels centered the position marked at both ends.
    Bottom: 1D spectra after subtracting the continuum in observed-frame (black line) and the observed Poisson noise (gray region).
    Orange line shows the fitting result to H$\beta$ and [O\,{\footnotesize III}]$\lambda\lambda$4959, 5007 emission lines (Section\,\ref{sec:line-fitting}). 
    Vertical hatched area shows the region contaminated by OH line remnants.}
    \label{fig:spec_J}
\end{figure*}

\begin{figure*}[h]
    \centering
    \begin{minipage}{0.22\textwidth}\centering\includegraphics[width=\linewidth]{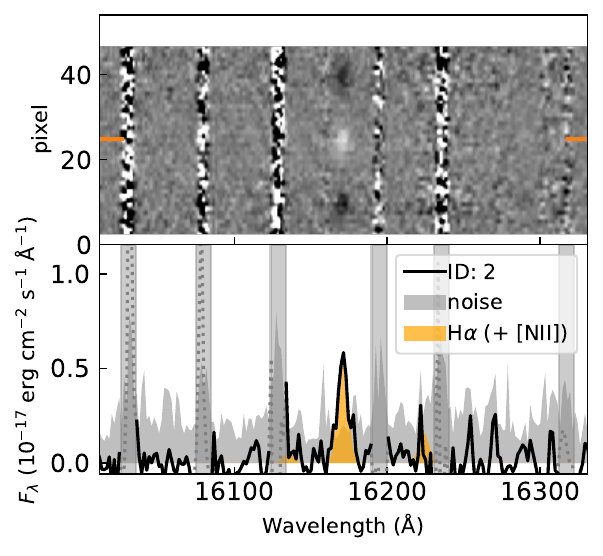}\end{minipage}\hfill
    \begin{minipage}{0.22\textwidth}\centering\includegraphics[width=\linewidth]{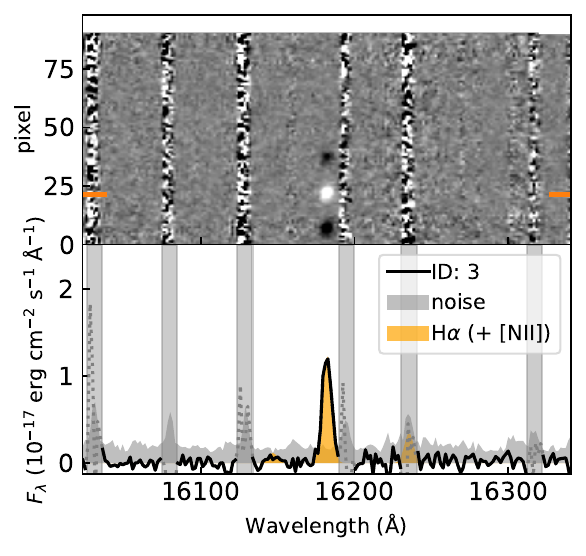}\end{minipage}\hfill
    \begin{minipage}{0.22\textwidth}\centering\includegraphics[width=\linewidth]{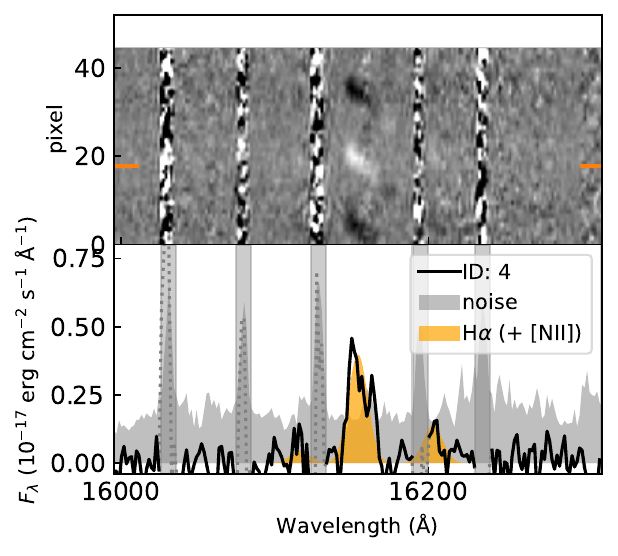}\end{minipage}\hfill
    \begin{minipage}{0.22\textwidth}\centering\includegraphics[width=\linewidth]{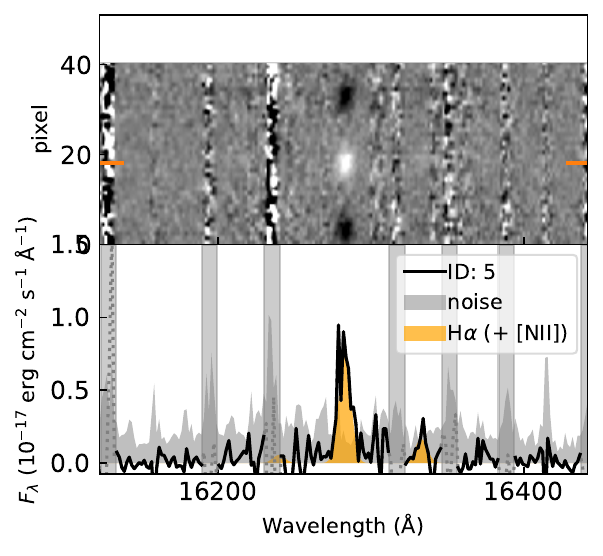}\end{minipage}
    \vspace{1ex}
    \begin{minipage}{0.22\textwidth}\centering\includegraphics[width=\linewidth]{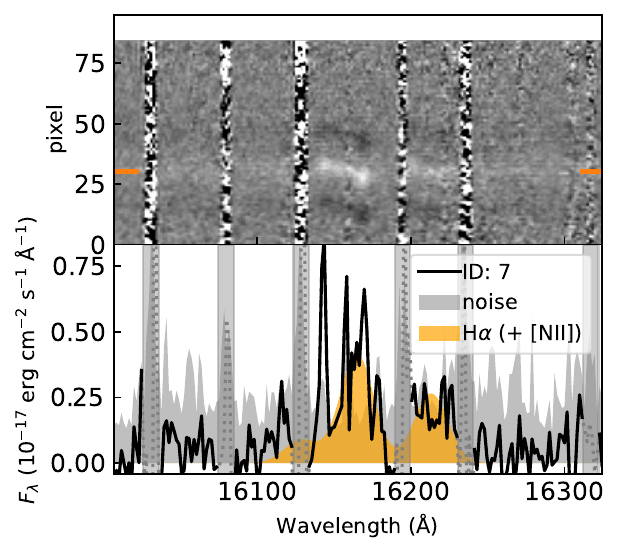}\end{minipage}\hfill
    \begin{minipage}{0.22\textwidth}\centering\includegraphics[width=\linewidth]{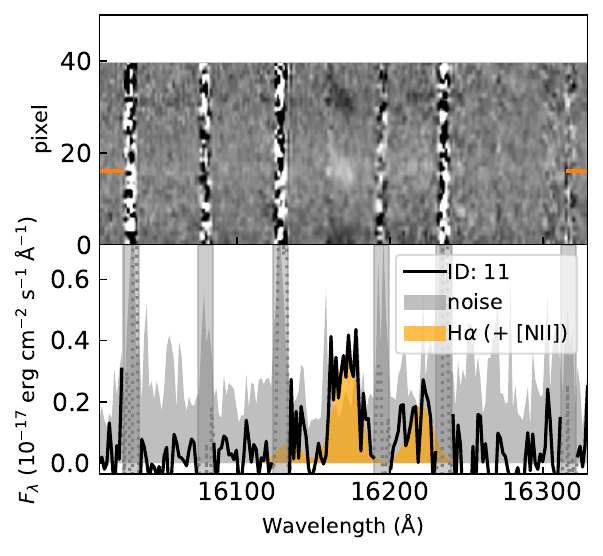}\end{minipage}\hfill
    \begin{minipage}{0.22\textwidth}\centering\includegraphics[width=\linewidth]{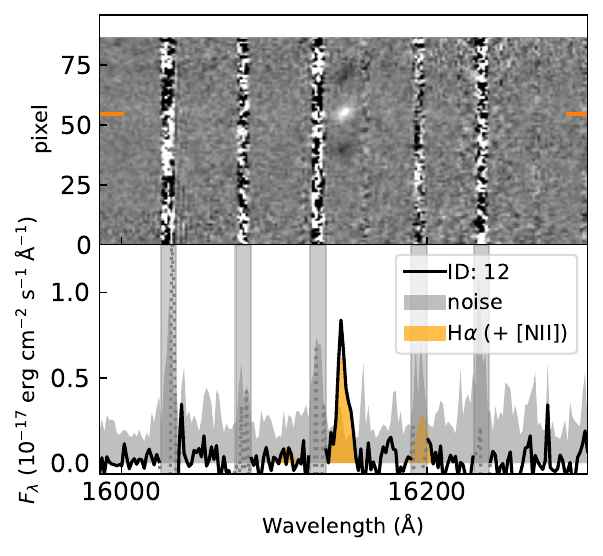}\end{minipage}\hfill
    \begin{minipage}{0.22\textwidth}\centering\includegraphics[width=\linewidth]{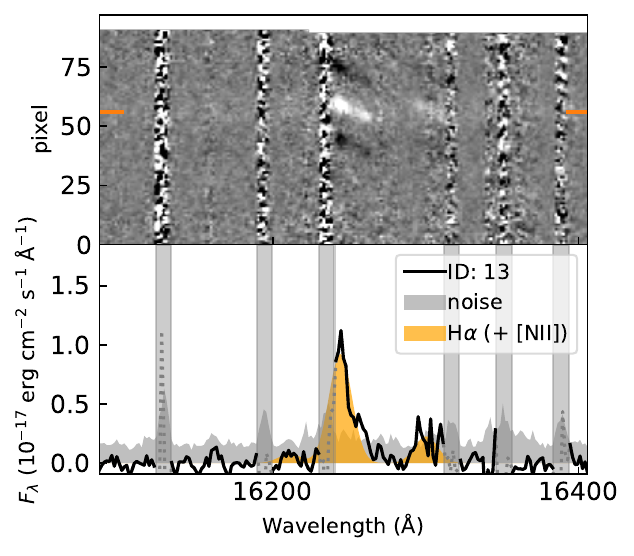}\end{minipage}
    \caption{Same as Figure \ref{fig:spec_J}, but for 2D \textit{H}-band spectra around the detected H$\alpha$ emission lines (Top), and 1D spectra and the fitting results to H$\alpha$ and [N\,{\footnotesize II}]$\lambda\lambda$6548, 6584 emission lines (Bottom).}
    \label{fig:spec_H}
\end{figure*}

\begin{figure*}[h]
    \centering
    \begin{minipage}{0.22\textwidth}\centering\includegraphics[width=\linewidth]{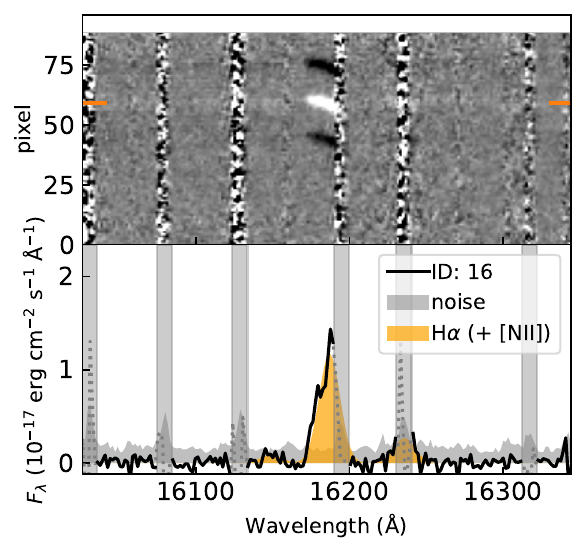}\end{minipage}\hfill
    \begin{minipage}{0.22\textwidth}\centering\includegraphics[width=\linewidth]{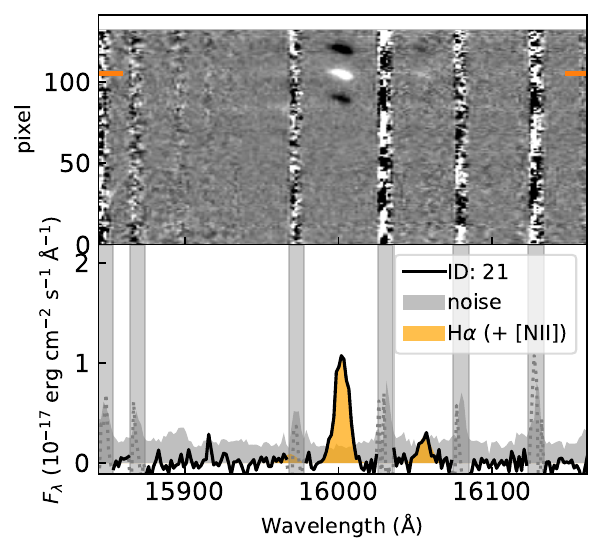}\end{minipage}\hfill
    \begin{minipage}{0.22\textwidth}\centering\includegraphics[width=\linewidth]{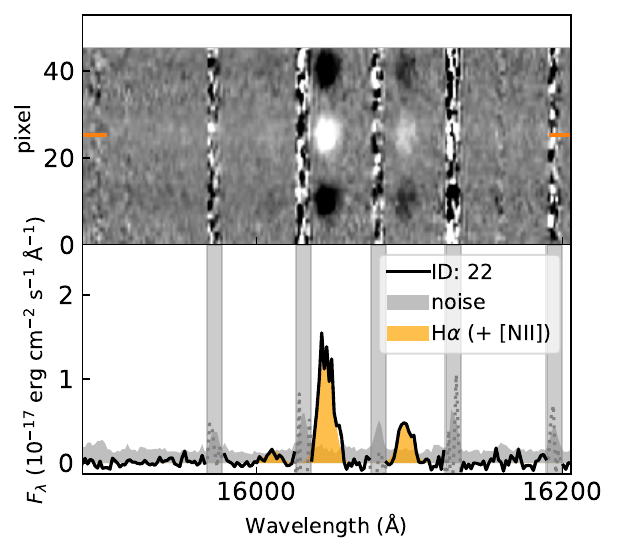}\end{minipage}\hfill
    \begin{minipage}{0.22\textwidth}\centering\includegraphics[width=\linewidth]{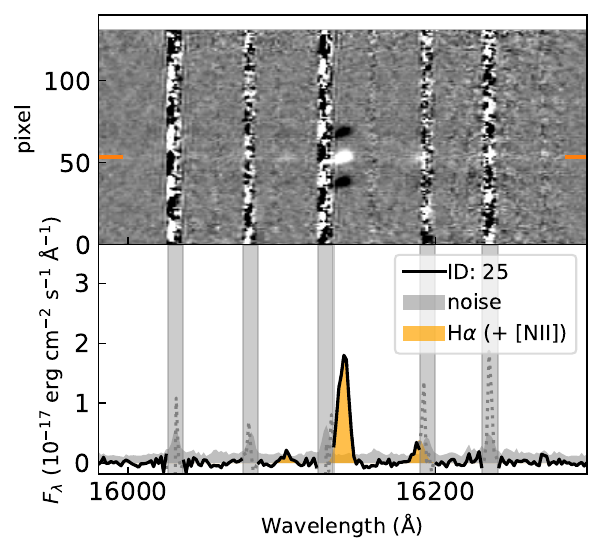}\end{minipage}
    \vspace{1ex}
    \begin{minipage}{0.22\textwidth}\centering\includegraphics[width=\linewidth]{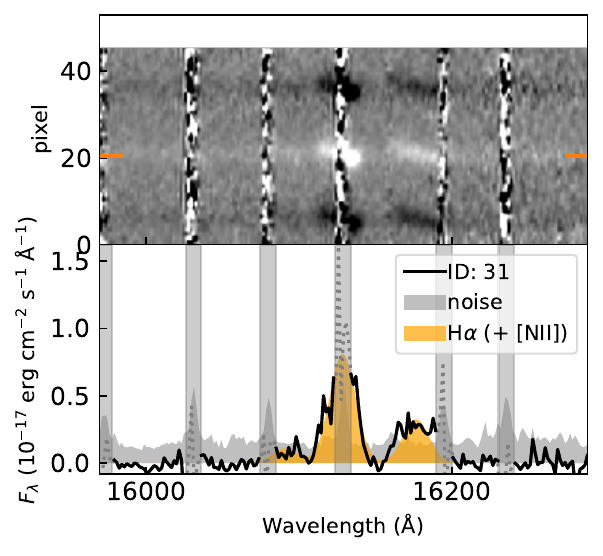}\end{minipage}\hfill
    \begin{minipage}{0.22\textwidth}\centering\includegraphics[width=\linewidth]{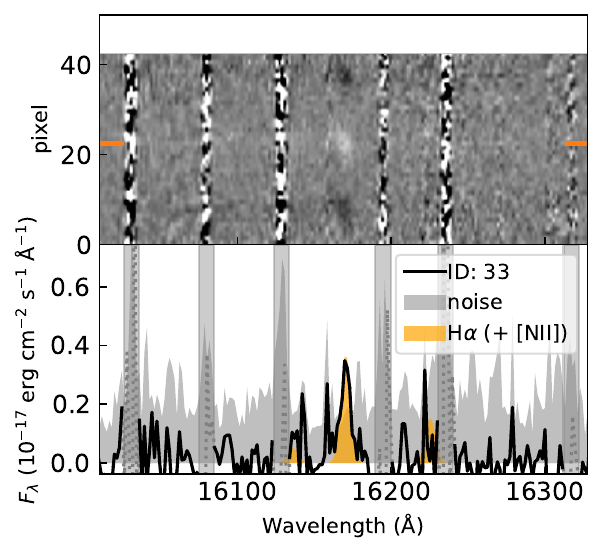}\end{minipage}\hfill
    \begin{minipage}{0.22\textwidth}\centering\includegraphics[width=\linewidth]{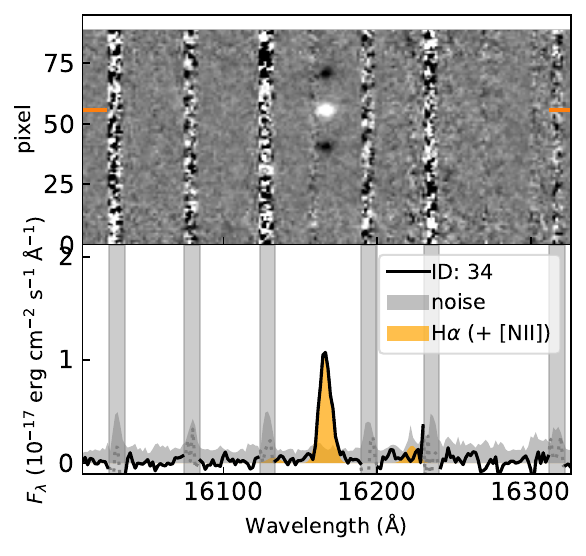}\end{minipage}\hfill
    \begin{minipage}{0.22\textwidth}\centering\includegraphics[width=\linewidth]{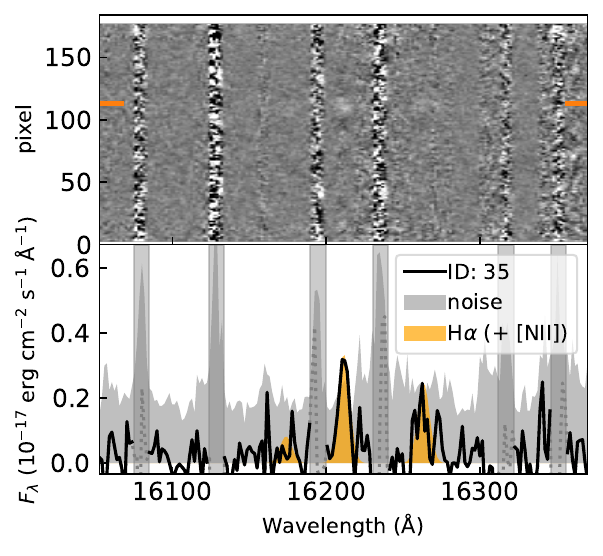}\end{minipage}
    \vspace{1ex}
    \begin{minipage}{0.22\textwidth}\centering\includegraphics[width=\linewidth]{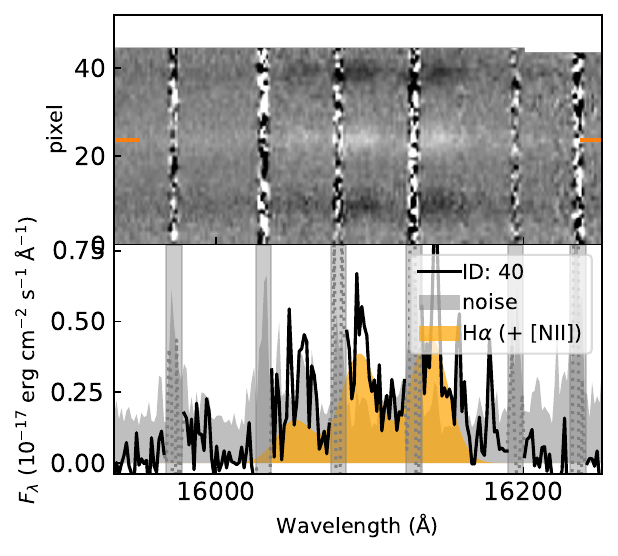}\end{minipage}\hfill
    \begin{minipage}{0.22\textwidth}\centering\includegraphics[width=\linewidth]{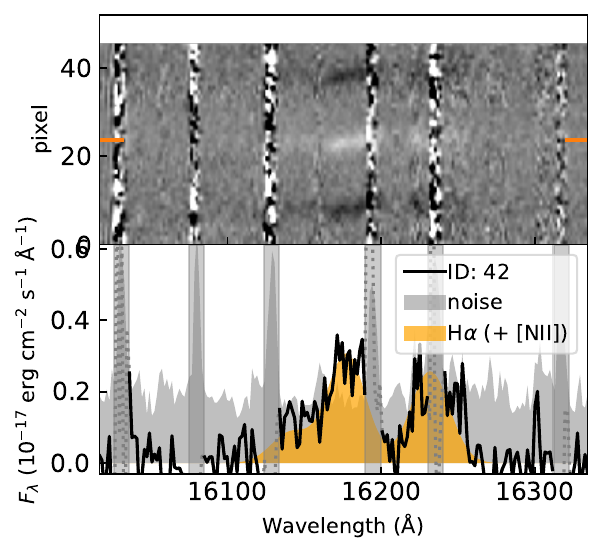}\end{minipage}\hfill
    \begin{minipage}{0.22\textwidth}\centering\includegraphics[width=\linewidth]{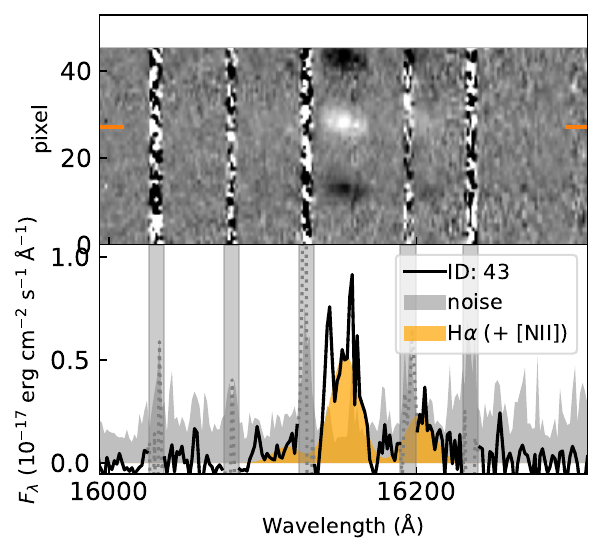}\end{minipage}\hfill
    \begin{minipage}{0.22\textwidth}\centering\includegraphics[width=\linewidth]{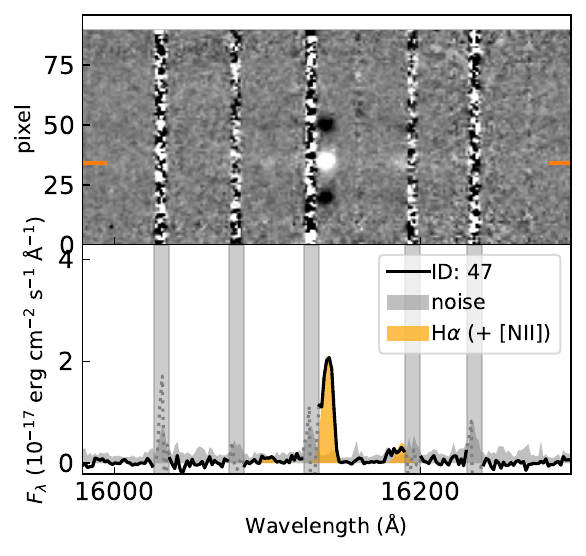}\end{minipage}
    \vspace{1ex}
    \begin{minipage}{0.22\textwidth}\centering\includegraphics[width=\linewidth]{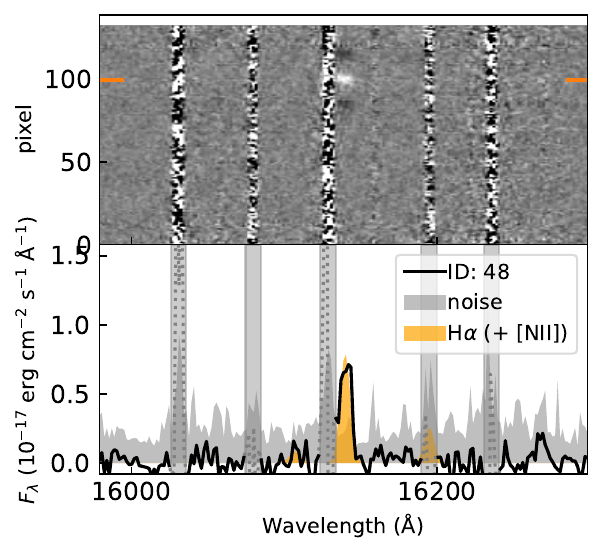}\end{minipage}\hfill
    \begin{minipage}{0.22\textwidth}\centering\includegraphics[width=\linewidth]{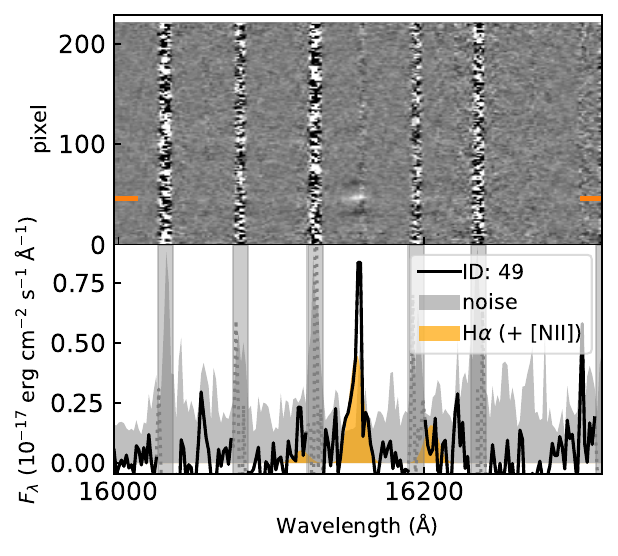}\end{minipage}\hfill
    \begin{minipage}{0.22\textwidth}\centering\includegraphics[width=\linewidth]{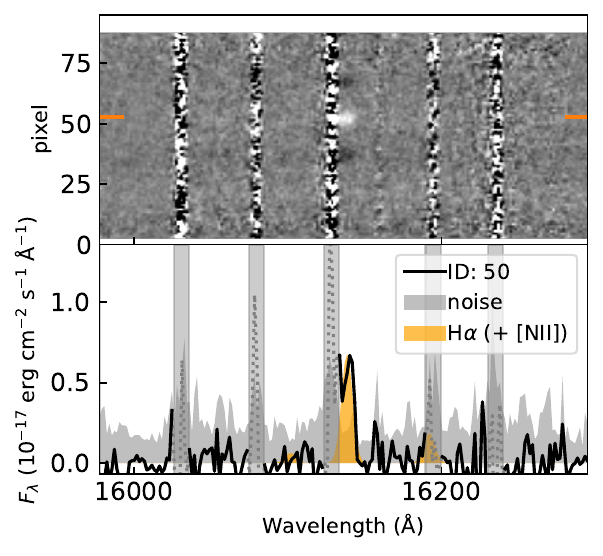}\end{minipage}\hfill
    \begin{minipage}{0.22\textwidth}\centering\rule{0pt}{0pt}\end{minipage}
    \caption{Figure \ref{fig:spec_H} continued.}
\end{figure*}

\clearpage
\section{Additional Figures} \restartappendixnumbering

\begin{figure}[h]
    \centering
    \includegraphics[width=0.5\linewidth]{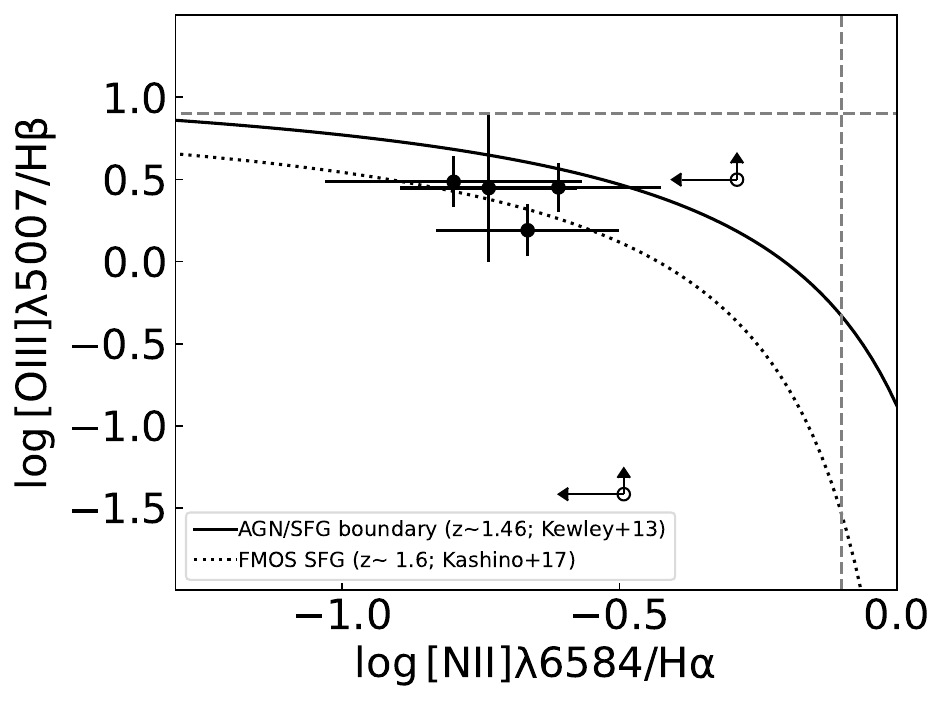}
    \caption{BPT diagram \citep{Baldwin_1981_classification_cn} for member galaxies (black points) with available H$\beta$ and [O\,{\footnotesize III}]$\lambda$5007 data by archival MOIRCS spectroscopy \citep{hayashi_properties_2011} or our MOSFIRE observation.
    Black solid line shows the boundary separating AGN from star-forming galaxies at $z\sim1.46$ \citep{Kewley_2013_cosmic_zw}.
    Gray dotted line shows the best fit to star-forming galaxies in FMOS-COSMOS survey \citep{kashino_fmos-cosmos_2017}.
    Gray dashed line shows another criteria of $\log [\mathrm{N\,{\footnotesize II}}]\lambda 6584/\mathrm{H}\alpha \geq -0.1$ or $\log [\mathrm{O\,{\footnotesize III}}]\lambda 5007/\mathrm{H\beta} \geq 0.9$ to identify AGN for the source with either line ratios available \citep{kashino_fmos-cosmos_2017}.}
    \label{fig:bpt}
\end{figure}

\begin{figure}[h]
    \centering
    \begin{minipage}{0.45\textwidth}\centering\includegraphics[width=\linewidth]{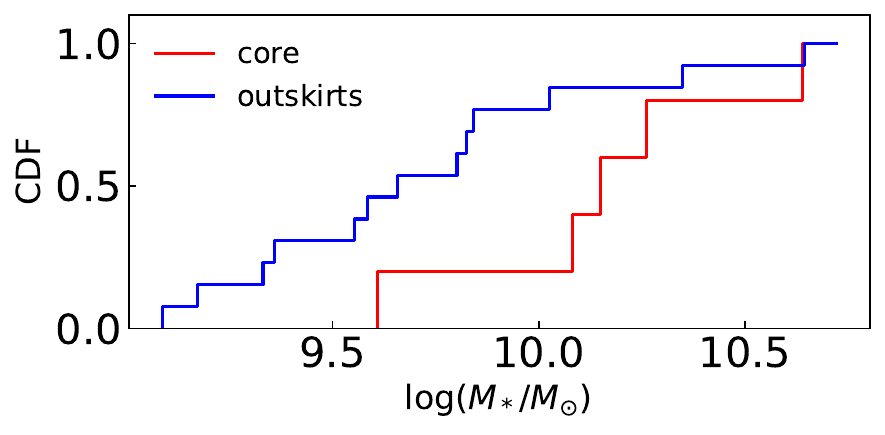}\end{minipage}\hfill
    \begin{minipage}{0.45\textwidth}\centering\includegraphics[width=\linewidth]{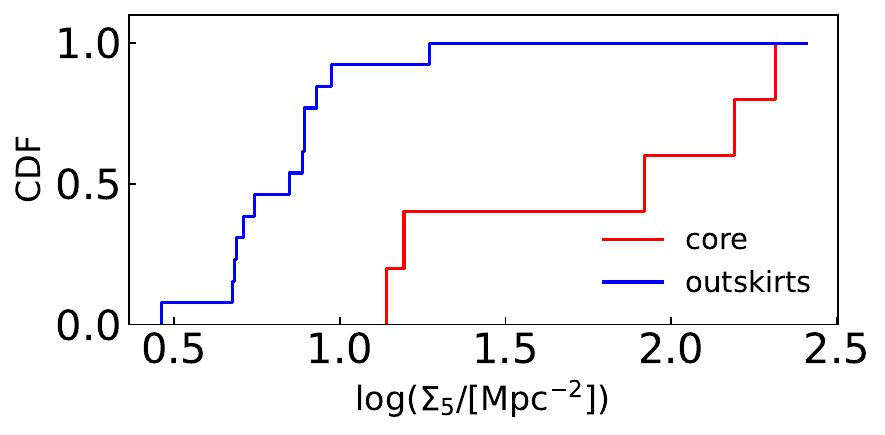}\end{minipage}\hfill
    \caption{Normalized cumulative distribution of stellar mass (Left) and local density $\Sigma_5$ (Right) for the galaxies used in the stacking analysis (Sec \ref{sec:mzr}).
    The red and blue line shows the distribution for the sources in the core and outskirts bins, respectively.}
    \label{fig:CDF}
\end{figure}

\bibliography{refs}{}
\bibliographystyle{aasjournal}
\end{document}